\documentclass{aa}  

\usepackage{graphicx}
\usepackage{natbib}
\bibpunct{(}{)}{;}{a}{}{,}
\usepackage{color}
\usepackage[dvipsnames]{xcolor}
\usepackage{txfonts}

\newcommand{\micron}{$\mu$m}
\newcommand{\Ks}{$K_{\rm{s}}$}
\newcommand{\msun}{M$_{\sun}$}
\newcommand{\lsun}{L$_{\sun}$}

\newcommand{\hii}{H{\sc ii}}

\newcommand{\brg}{Br$\gamma$}
\newcommand{\kms}{km~s$^{-1}$}
\newcommand{\hei}{\ion{He}{I}}
\newcommand{\heii}{\ion{He}{ii}}
\newcommand{\niii}{\ion{N}{iii}}

\newcommand{\teff}{$T_{eff}$}
\newcommand{\ak}{$A_{\mathrm{K_{s}}}$}

\begin{document} 

   \title{Near-infrared spectroscopy of the massive stellar population of W51: evidence for multi-seeded star formation.
   \thanks{Based on data acquired using the Large Binocular Telescope (LBT). The LBT is an international collaboration among institutions in Germany, Italy and the United States. LBT Corporation partners are: LBT Beteiligungsgesellschaft, Germany, representing the Max Planck Society, the Astrophysical Institute Potsdam, and Heidelberg University; Instituto Nazionale di Astrofisica, Italy; The University of Arizona on behalf of the Arizona university system; The Ohio State University, and The Research Corporation, on behalf of The University of Notre Dame, University of Minnesota and University of Virginia.},\thanks{Based on observations made with ESO Telescopes at the La Silla Paranal Observatory under programme ID 079.C-0248(A).}}

   \subtitle{}
   \titlerunning{Near-infrared spectroscopy of the massive stellar population of W51}

   \author{A. Bik\inst{1}\fnmsep\inst{2}
          \and
          Th. Henning\inst{2}
          \and
   S. -W. Wu\inst{2}\fnmsep\thanks{International Max Planck Research School for Astronomy and Cosmic Physics at the University of Heidelberg (IMPRS-HD)}
          \and
          M. Zhang\inst{2}
          \and
          W. Brandner\inst{2}
          \and
          A. Pasquali\inst{3}
          \and
	A. Stolte\inst{4}
          }

   \institute{Department of Astronomy, Stockholm University, AlbaNova University Centre, 106 91 Stockholm, Sweden\\
    \email{arjan.bik@astro.su.se}
             \and
                Max-Planck-Institut f\"{u}r Astronomie, K\"{o}nigstuhl 17, 69117 Heidelberg, Germany
                         \and
             Astronomisches Rechen-Institut, Zentrum f\"ur Astronomie der Universit\"at Heidelberg, M\"onchhofstr. 12 - 14, 69120 Heidelberg, Germany        
             \and
             Argelander Institut f\"ur Astronomie, Auf dem H\"ugel 71, 53121 Bonn, Germany
             }

   \date{Received ; accepted }

 
  \abstract
   {The interplay between the formation of stars, stellar feedback and cloud properties strongly influences the star formation history of giant molecular clouds. The formation of massive stars leads to a variety of stellar clusters, ranging from low stellar density OB associations to dense, gravitationally bound starburst clusters. }
   {We aimed at identifying the massive stellar content and reconstructing the star formation history of the W51 giant molecular cloud.}
   {We performed near-infrared imaging and K-band spectroscopy of the massive stars in W51. We analyzed the stellar populations using colour-magnitude and colour-colour diagrams and compared the properties of the spectroscopically identified stars with stellar evolution models.}
   {We derived the ages of the different sub-clusters in W51 and, based on our spectroscopy derived an age for W51 of 3 Myrs or less. The age of the P Cygni star LS1 and the presence of two still forming proto-clusters suggests that the star formation history of W51 is more complex than a single burst.}
   {We did not find evidence for triggered star formation and we concluded that the star formation in W51 is multi seeded. We finally concluded that W51 is a OB association where different sub-clusters form over a time span of at least 3-5 Myrs.}


   \maketitle
%

\section{Introduction}

Stars form deeply embedded in giant molecular clouds (GMCs), and especially the formation process of massive stars is still a subject of debate \citep[e.g.][]{Zinnecker07,Tan14}.  What has become clear is that stars do not form isolated but in stellar clusters and associations \citep[e.g.][]{Lada03,Krumholz14problems}. This means that the forming massive stars interact with each other by means of stellar feedback, where the first generation of star formation affects future generations of star formation in the same cloud \citep[e.g.][]{Krumholz14PPVI}.

The star formation process is hierarchical, driven by the self similar distribution of the dense gas in the molecular clouds \citep[e.g.][]{Elmegreen08,Elmegreen10}.  This results in different observed properties of star forming regions. Young star clusters with high stellar densities are  in general found to be gravitationally bound \citep[e.g.][]{Stolte04,Stolte06,Ascenso07,Stolte08,Brandner08,Rochau10,Harfst10,Gennaro11,Hussmann12}. On the other side of the hierarchical spectrum  extended, low-stellar density OB associations are found \citep[e.g.][]{Blaauw91,deZeeuw99,Comeron02,Preibisch11}. 
The star formation histories of these star forming regions suggest that the star clusters such as Westerlund 1 and NGC 3606 consist of a single age population \citep[e.g.][]{Clark05,Negueruela10,Kudryavtseva12}, while less compact clusters such as the ONC \citep{Palla99,DaRioOrion10}, W3 Main \citep{Feigelson08,Bik12} show evidence of an age spread of several Myrs.  In larger OB associations, such as the Orion OB association and Cyg OB2 much large age differences between OB subgroups are measured. In the case of Orion an age difference of at least 12 Myrs \citep{Brown94,Bally08} is present, while analysis of the massive stars in Cyg OB2 shows that star formation has been going on for at least 10 Myrs \citep{Comeron12,Berlanas18}. Also older populations have been found associated with starburst clusters such as NGC3603, where low-mass pre-main-sequence (PMS) stars up to ages of 10 Myrs have been identified \citep{Beccari10} as well as two evolved supergiants with an age of 4 Myrs \citep{Melena08}.  The observation that the age difference between the stars increases with the size of the observed structure can be explained by the hypothesis that star formation happens in a crossing time \citep{Elmegreen00}. 

Kinematical studies based on proper motions  confirm this fundamental difference between OB associations and starburst clusters.  The starburst clusters consist of a single stellar population and their kinematics show typically a low velocity dispersion \citep{Stolte08,Rochau10,Hussmann12}. OB associations on the other hand show  more complicated kinematics. \citet{Wright16} showed, based on multiple epoch imaging, that the  Cyg OB2 association  consists of multiple subgroups moving at different velocities and directions. The kinematics suggest that the stars are formed in-situ as a result of multiple small scale star formation events.  This result is inconsistent with OB associations being expanding clusters \citep[e.g.][]{Baumgardt07}. With  \emph{Gaia} \citep{Gaia16}, the in-situ star formation scenario  has been confirmed for other OB associations \citep{Wright18,Ward18,Kuhn18}. 

The differences in morphology and star formation history will  have a strong impact on how the feedback of the forming stars act on the molecular cloud \citep[e.g.][]{Rey-Raposo17,Kim18}. Feedback influences future generations of stars by triggering new episodes of star formation or destroying the molecular cloud.  Therefore a detailed understanding of the physical processes and stellar content in different GMC environments is important to understand how star formation progresses in different environments.


With the LOBSTAR project we are aiming at studying the stellar content of  some of the most luminous HII regions in our Galaxy in order to trace back their formation history. We obtain high signal-to-noise ratio (S/N) K-band spectra of the massive stars and deep near-infrared imaging of these regions. In \citet{Bik12,Bik14} we presented evidence of an age spread of a few Myrs in the stellar population of W3Main and that the high UV flux of the massive stars affects the circumstellar disks around the young stars. In W49, \citet{Wu14} discovered one of the few known very massive stars (VMS) with a photometrically estimated mass between 100 and 180 \msun. \citet{Wu16} performed a spectral analysis of 14 O-type stars in W49 and concluded that the cluster has an age of $\sim$ 1.5 Myrs. 

This paper focusses on the last of the observed regions:  W51. The W51 GMC complex is an elongated molecular cloud complex, located in the Sagittarius spiral arm at a distance of 5.41$^{+0.31}_{-0.28}$ kpc \citep{Sato10}. W51 is one of the few GMC complexes in our galaxy with a mass above 10$^6$ \msun\ \citep{Carpenter98} and spans a region on the sky of $45' \times 50'$, corresponding to 70 $\times$ 80 pc.  W51 consists of three parts; W51A is located in the north-east (high galactic longitude), W51B is located in the south-west (low galactic longitude) and W51C, a supernova remnant \citep{Koo95,Koo97,Brogan13}, is located   towards the south-east  of the complex.  An observational review of the W51 complex can be found in   \citet{Ginsburg_SF17}.

The region of most active star formation in W51 is the W51A cloud. This cloud has a much higher fraction of dense gas than W51B \citep{Ginsburg15} and hosts two of the only few known proto-clusters able to form a 10$^4$ \msun\ or higher  star cluster \citep{Ginsburg12}.  W51A was originally  identified as a luminous radio source \citep{Westerhout58}. High resolution VLA observations have resolved W51A in many compact \hii\ regions \citep{Martin72,Scott78,Mehringer94,Ginsburg16}. Additionally several HII regions are found in W51B as well as radio emission towards the supernova remnant W51C \citep{Koo97}. W51 has been studied in  molecular transitions such as CO \citep{Carpenter98,Kang10,Fujita17}, \ion{H}{i} \citep{Koo97} and several other more complex molecules.

%

\begin{table*}[!t]
\caption{SOFI Observation of W51: Log of Observations}
\centering
\begin{tabular}{crccccccccc}
\hline\hline 
Field & HII region & $\alpha$ (J2000)&$\delta$ (J2000) &Observing date & \multicolumn{3}{c}{Exposure time(s)} & \multicolumn{3}{c}{Seeing(\arcsec)}\\
\cline{6-8} \cline{9-11}\\
 &&(h m s)&($^\circ$\ \arcmin\  \arcsec)&(YYYY-MM-DD)&J&H&K$_s$&J&H&K$_s$ \\
\hline
I & G49.58-00.38&19:23:52.4 & 14:35:27&2007-08-30 &288 &288 &288&1.05&1.16 &1.29\\
II &G49.5-0.4&19:23:43.5 & 14:31:10&2007-09-28 &288 &288 &288&1.07&1.04&0.95 \\
III &G49.2-0.3&19:23:01.9 & 14:16:17&2007-09-30 &288 &288 &288&1.00&0.88&1.08 \\
IV & G48.9-0.3&19:22:13.7 & 14:02:09&2007-08-27 &288 &288 &230&1.00&1.26&1.13\\
\hline\label{tab:obslog}
\end{tabular}
\end{table*}

The stellar populations of the entire W51 GMC are analyzed in detail in several near-infrared imaging studies. Most observations focus on the central regions of  the W51A cloud where \citet{Neugebauer69} and \citet{WynnWilliams74} detected several bright infrared sources, which were resolved in small clusters in higher spatial resolution  imaging \citep{Goldader94}. 
\citet{Okumura00} studied the  stellar content of the entire W51A  cloud using near-infrared $JHK'$ and \brg\ imaging. They identified several candidate OB stars and presented evidence for a top heavy initial mass function (IMF). Four subgroups have been found, ranging in age between 1 and 2.2 Myrs.  \citet{Kumar04} observed 6 fields in the W51 GMC, including W51B, centered on the suspected location of embedded clusters based on 21-cm radio emission \citep{Koo97}, MSX and 2MASS data.  \citet{Figueredo08} obtained low resolution $K$-band spectra for five objects in the region of W51A, and could classify four of them as O-type stars. High resolution $K$-band spectra of two massive stars (associated with IRS2E and IRS2W, respectively) were obtained by \citet{Barbosa08} in W51d, one of the compact \hii\ regions in W51A. IRS2E was identified as a young stellar object (YSO), and IRS2W -- which turns out to be the main ionizing source of W51d  -- was classified as an early O-type star \citep{Barbosa08}.

Additional to the candidate OB stars, \citet{Okumura00} identified two stars which are more luminous than the typical OB stars in W51. Both sources (LS1 and LS2) show an emission line spectrum. Based on the observed emission lines of \brg, \hei\ and \ion{Mg}{II} in the K-band spectrum of LS1, the authors concluded that LS1 is a very massive P Cygni-type supergiant. Detailed modeling of a higher spectral resolution K-band spectrum by \citet{Clark09} confirmed this conclusion, although the object is significantly less massive ($\sim$40 \msun) than the mass reported by \citet{Okumura00}. Spitzer imaging shows that the source is surrounded by a wind blown bubble. \citet{Clark09} showed that the spectral identification as P Cygni supergiant places the objects at the distance of W51. 
Under the assumption that LS1 is part of the W51 cloud, \citet{Clark09} showed that the age of the LS1 suggests that star formation has been active in W51 for at least 3 Myrs. LS2 shows \brg\ emission and could be either a B[e] supergiant or a young star surrounded by a circimstellar disk \citep{Okumura00}.

The more embedded stellar content is traced with Spitzer, Herschel  and SOFIA observations. Spitzer observations reveal hundreds of YSOs associated with the W51 cloud \citep{Saral17} distributed in several sub-clusters.  They identified 17 massive YSO candidates and found that their locations correlate with the \hii\ regions detected in the radio. \citet{Eden18} presented Herschel and JCMT observations revealing a large sample of mm point sources in both W51 and W49A. They showed that the luminosity distribution of embedded sources in  W51 and W49A are different. W51 has a luminosity function which is similar to the galactic average, while W49A shows an excess in bright sources, suggesting that the physical conditions in the central region of the W49A cloud could be different than in W51 and represents a more extreme mode of star formation.
 Using SOFIA imaging at 20 and 37 \micron\ \citet{Lim19} studied the central areas of W51A at resolution of 2.2\arcsec (20 \micron) to  3.0\arcsec (37 micron). They detected the dusty cocoons around the \hii\ regions observed in the radio, showing that many \hii\ regions are still deeply embedded.  By combining the SOFIA imaging with Spitzer and Herschel, they identified 41 candidate massive YSOs. This demonstrates the very young nature of W51A.

The \hii\ regions hosting the most important stellar clusters in W51A are called G49.5-0.4 and G49.4-0.3 of which the latter is hosting the two proto-cluster regions W51IRS2 and W51Main \citep{Ginsburg12}. W51B hosts G49.2-0.3, G49.1-0.4, G49.0-0.3 and G48.9-0.3 \citep{Koo97,Kumar04}, all harbouring embedded clusters. In Appendix \ref{sec:subclusters} we describe the properties of each \hii\ region in more detail.

In this paper, we present near-infrared multi-object (MOS) spectroscopy with LBT/LUCI and NTT/SOFI near-infrared imaging  of the stellar population of the clusters inside four \hii\ regions: G49.58-00.38 and G49.5-0.4 in W51A and  G49.2-0.3 and G48.9-0.3 and W51B. The reduction of the photometric and spectroscopic data are presented in Sect.~\ref{W51:reduction}. In Sect.~\ref{sec:results}, we classify the spectra of the massive stars, derive their properties and  place them in a Hertzsprung-Russell diagram  (HRD).  In Sect.~\ref{W51:discussion}, we discuss the cluster properties and the star formation history in more detail.  Conclusions are drawn in Sect. \ref{sec:conclusions}.

\begin{table*}[!h]
\caption{LUCI multi-object-spectroscopy: Log of Observations}
\centering
\begin{tabular}{c|cccl}
\hline\hline 
Field & mask name &Observing date & Exposure time &Comment\\
 & & (YYYY-MM-DD) & (s) & \\ \hline
I 	& OBS5&2012-06-28& 3600 &					\\
II	& OBS1&2012-05-28& 3600& standard star shifted w.r.t. flat-field		\\
	& OBS2&2012-05-30& 3600&					\\
	& OBS3&2012-05-30& 3600&					\\
	& OBS4&2012-06-25& 3600&					\\
III	&OBS6&2012-10-08& 3600& science and standard star shifted w.r.t. flat-field \\
	&OBS9&2012-10-11& 3600&					\\
IV	&OBS7&2012-10-10& 3600&science and standard star shifted w.r.t. flat-field\\
	&OBS8&2012-10-11& 3600&science and standard star shifted w.r.t. flat-field\\
\hline\label{tab:spec}

\end{tabular}
\end{table*}

\section{Observation and data reduction}
\label{W51:reduction}


\subsection{Observations}
The imaging observations  of the W51 GMC were conducted in 2007 between  August 27 and September 30, (PID: 079.C-0248(A), PI: J. S. Clark) with the  spectrograph and imaging camera SOFI \citep{Moorwood98SOFI}, mounted on the New Technology Telescope (NTT) on La Silla, Chile.
The instrument  is equipped with a Hawaii HgCdTe 1024 $\times$ 1024 array, yielding a field of view (FOV) of $\sim$ 4\arcmin.9 $\times$ 4\arcmin.9 with a plate scale of 0\arcsec.288 pixel$^{-1}$ in its large field imaging mode. 

In total, 8 fields were observed in $J$, $H$, and \Ks-bands, however in this paper we only use 4 of the observed fields for which multi-object spectroscopy was acquired (Tab. \ref{tab:obslog}).   Using dithered observations, each field was typically imaged ten times with individual exposure times of 28.8s, with a detector integration time (DIT) of 1.2 s repeated 24 times (NDIT) in all three bands. Table \ref{tab:obslog} lists the centre coordinates, the observing date, the total integration time, and the average seeing for each field.

The spectroscopic targets were selected by constructing colour-magnitude diagrams (CMDs) from the SOFI photometry data of each cluster.  Similar to \citet{Okumura00}, a comparison with stellar evolutionary tracks revealed the candidate OB stars in each region. We selected reddened sources which are bright enough to be reddened OB stars at the distance of W51 and have a clear visual association on the near-infrared images with the stellar cluster or the \hii\ region. 
 These candidate OB stars make the high priority target list. However, due to the nature of the MOS observations, not all stars could be placed in a slit, as that would cause overlapping slits. Most regions were observed with more than 1 mask, but still several good candidates were left out. As the clusters are typically smaller than the LUCI field of view, the open spaces in the mask were filled up with additional, lower priority, sources, not obviously related to the clusters.

The $K$-band spectroscopic data of the candidate massive stars in W51 were observed with  LUCI1 \citep{Ageorges10,Seifert10} mounted on the Large Binocular Telescope \citep[LBT,][]{Hill06} on Mount Graham, USA. We used LUCI1 in  MOS mode \citep{Buschkamp10} using the  210\_zJHK grating with the N1.8 camera, providing a 0\farcs25 pixel scale.  This results in K-band spectra with a  wavelength coverage of  $\Delta\lambda$=0.328 \micron. The central wavelength of the spectra depends on the position in the field of view. 

The observations were carried out between  May 28 and  October 11 2012. The observations were split in 12 frames with a DIT of 60 s and NDIT of 5 per frame. In total 9 masks were observed covering the four different fields imaged with SOFI (see Tab. \ref{tab:spec} for a summary).  The slits  in the masks have a width of 1\arcsec, resulting in a spectral resolution of R=4000, and a length of 10\arcsec\ in order to enable good sky subtraction. The frames were observed with a nodding offset of 5\arcsec.
 Per observation, two telluric standard stars were observed (before: Hip 90337, B9V, and after: Hip 104320, B3V)  in each slit to ensure that wavelength coverage and other observational conditions were the same as for the scientific observations, resulting in the best possible correction for the telluric absorption lines. 

Due to technical issues with the telescope, the science and the standard star data for three observations (OBS6, OBS7 and OBS8) show an offset in the spatial direction between each frame. This means that only parts of the slits are covered by the flat-field. In the course of the data reduction process we  only selected those spectra which overlapped with the flat-fields. Also the standard star observation of OBS1 was shifted with respect to the flat-field.

\begin{figure*}[!th]
   \includegraphics[width=\hsize]{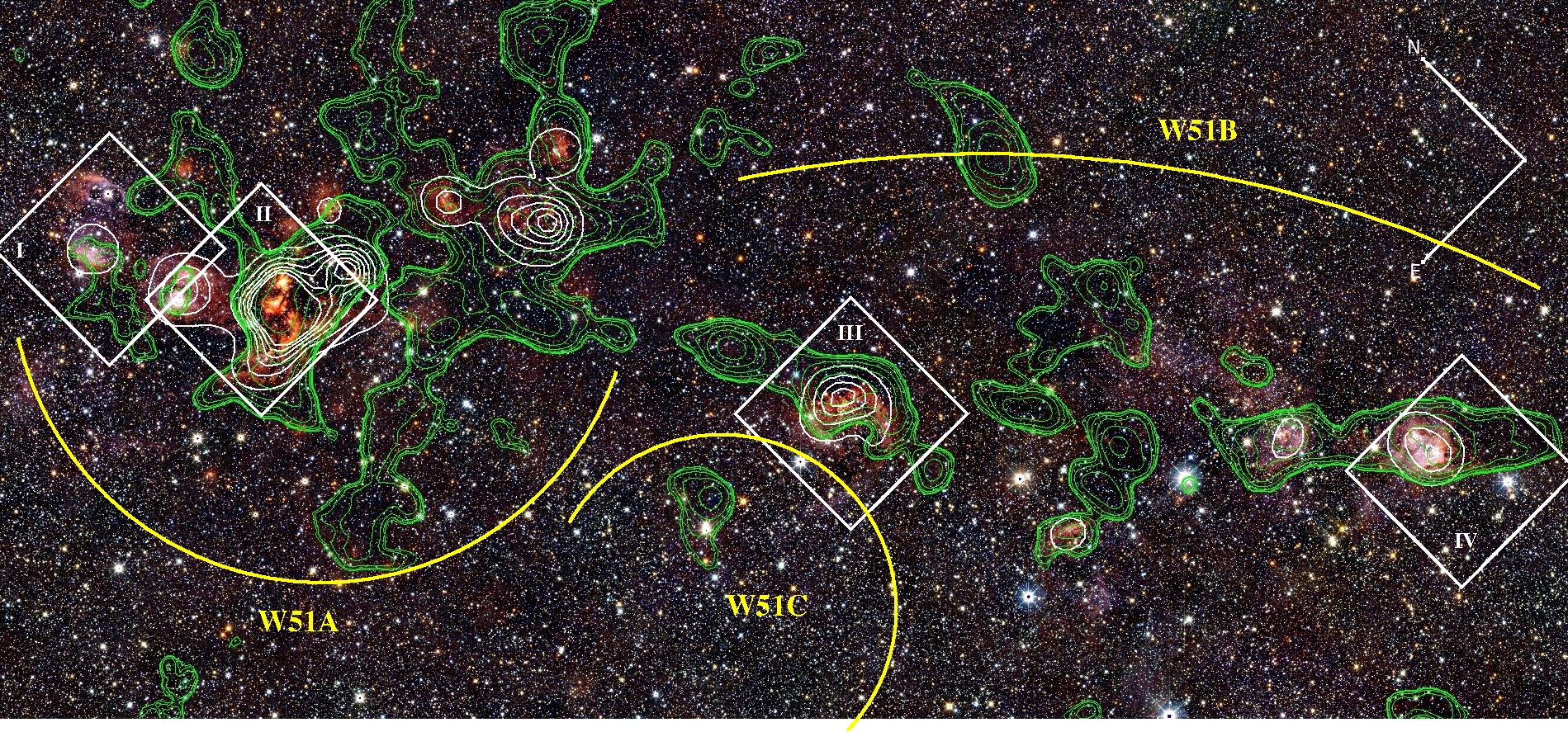}
      \caption{Three-colour  $JH$\Ks-band image from 2MASS of W51 (20\arcmin $\times$ 40\arcmin), overlaid by 1.1-mm dust continuum contour from the Bolocam survey \citep{Aguirre11} in green and the 1.4 GHz NRAO VLA Sky Survey \citep[NVSS,][]{Condon98} contours in white. The four white boxes mark the areas where photometry and spectroscopy are obtained. From  left to right, field I and II cover  \hii\ region G49.5-0.4, field III G49.2-0.3 and field IV covers G48.9-0.3. The yellow arcs mark the location and extent of the 3 subregions in W51: W51A, W51B and W51C used throughout the paper.}
    \label{fig:largescale}
\end{figure*}

\subsection{Data reduction}

\subsubsection{Imaging}

The imaging data reduction is performed using an IDL pipeline based on  \textit{eclipse} routines \citep{Devillard01} and improved by \citet{Gennaro12}.  The raw data were processed with dark subtraction, flat-fielding, cross-talk removal, and sky subtraction. We also removed the geometrical distortion on the dithered frames using a geometric distortion solution. Finally we combined the dithered frames in each filter for each field using the \textit{eclipse jitter} task. The absolute astrometry is performed on the coadded frames by comparing the image with the 2MASS reference catalog \citep{Skrutskie06} using the IRAF \textit{ccmap} and \textit{cctran} tasks. We used the UKIDSS\footnote{The UKIDSS project is defined in \citet{Lawrence07}. UKIDSS uses the UKIRT Wide Field Camera (WFCAM; \citet{Casali07}) and a photometric system described in \citet{Hewett06}. The pipeline processing and science archive are described in \citet{Irwin08} and \citet{Hambly08}. We have used data from the 7th data release, which is described in detail in \citet{Warren07}.} GPS \citep{Lucas08} catalog to estimate the accuracy of our astrometry.  We found that the rms values for all fields are below 0\farcs2. 

\begin{figure*}[!t]
 \includegraphics[width=\hsize]{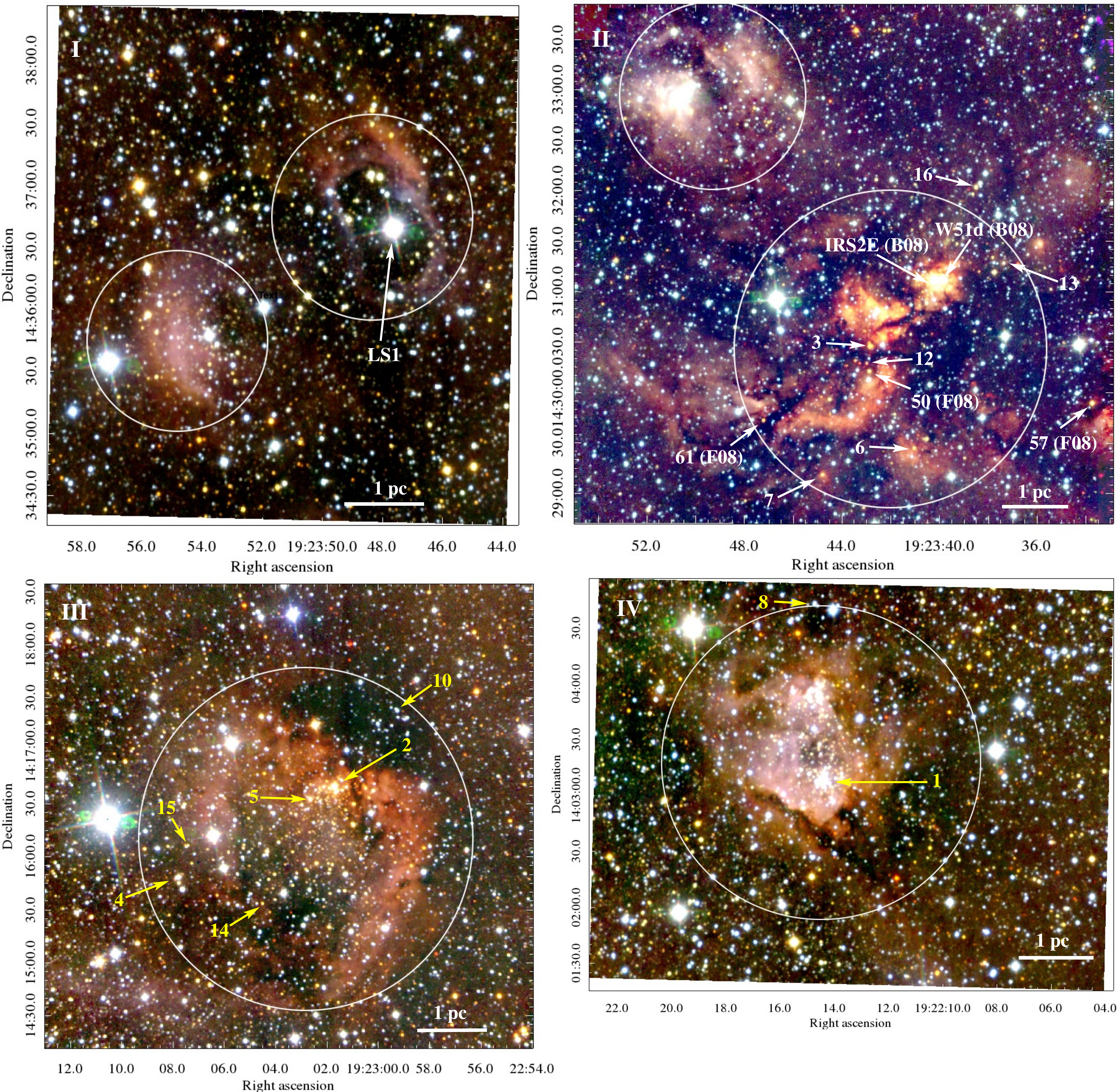}
      \caption{$JH$\Ks\ composite colour image from the SOFI data of the 4  observed fields  in W51, zoomed in on the clusters, from left to right  field I, field II, field III and field IV.  The red, green and blue channels are mapped to the \Ks, $H$ and $J$, respectively. The regions selected for the photometric analysis are indicated with white circles. The spectroscopically classified massive stars in each field are indicated with their ID numbers.
In field II, also the positions of the  O stars and massive YSO from \citet{Figueredo08} and \citet{Barbosa08} are shown.
      }
    \label{fig:W51subregion}
\end{figure*}

Photometry was performed on the reduced frames using the IRAF \texttt{DAOPHOT} package \citep{Stetson87}. Point spread function (PSF) fitting photometry was conducted on the co-added images in each filter of each field. Photometric zero points and colour terms were calculated by a comparison of the instrumental magnitudes of relatively isolated, unsaturated bright sources with the counterparts in the UKIDSS catalog. By comparing $\sim$12,500 sources detected in both the SOFI and the UKIDSS observations, we found that the photometric reliability in all bands is $\sim$0.04-0.14 mag, depending on the source brightness. In this paper we only used the sources with a photometric error of 0.2 mag or less in all three bands.

\subsubsection{Spectroscopy}
The LUCI spectra were reduced with a modified version of \emph{lucired}, using the same method as described in \citet{Bik12} and \citet{Wu16}. First, the frames were corrected for tilt of the slits and geometric distortion by a spectroscopic sieve mask and an imaging pinhole mask respectively. Afterwards, dark subtraction and division by a normalized flat-field were performed. Individual slits were extracted from the spectroscopic frames and wavelength calibration was performed for each slit using a  Argon and Neon wavelength calibration frame. 

The frames were corrected for the sky background by subtracting two frames adjacent in time at different nodding positions. In some cases this did not result in a good subtraction of the sky emission lines and the procedure of \citet{Davies07} was used. This procedure corrects for variations in the OH emission lines by fitting the individual transitions.  One dimensional spectra were extracted using \emph{doslit} under IRAF and all the spectra of the same source averaged to one final spectrum.  

Due to the offsets in the spatial direction of the observations from OBS6, OBS7 and OBS8, the science spectra do not always spatially overlap with the flat-fields. This severely degraded the quality of the spectra. Therefore, we selected only those parts of the slits where there was spatial overlap with the flat-field observations. This means that some slits could not be used and in others we lost up to 50 \% of the recorded spectra, resulting in lower S/N spectra. 

Before performing the telluric correction, the \brg\ absorption line in the standard star spectrum was removed by fitting the line with a Lorentzian profile. After that  the science spectra were corrected with the telluric standard stars taken before or after the science observations using the IRAF task \emph{telluric}.   The science spectra with the least residuals were selected as final results. The standard star observations of OBS6, OBS7 and OBS8 were also affected by the shifted location of the spectra with respect to the flat-field, thus could not be used for the telluric correction.  This is also the case for OBS1 where only the standard star observations were affected by a shift.
To remove the telluric lines from the science observations of OBS1, 6, 7 and 8 for telluric lines we used MOLECFIT \citep{Smette15,Kausch15}. MOLECFIT fits synthetic transmission spectra to the observed spectra taking into account the atmospheric conditions during the observations.

For this procedure, first the atmospheric conditions on the day of the observations for Mount Graham were downloaded\footnote{\url{http://www.ready.noaa.gov/READYamet.php}}. After that the spectra were fitted with transmission models including H$_{2}$O, CO$_{2}$, N$_{2}$O, CO and CH$_{4}$. The best matching spectral resolution for the models and the wavelength shift were calculated first by fitting a small wavelength range without varying the abundances. After those values are determined, the entire wavelength range was fitted with a fixed spectral resolution. 
This two step procedure gave the best resulting telluric correction.

For OBS1slit17 (\#16) the abundance fit was performed in two stages. The region redwards of 2.29 \micron\ in the spectrum of \#16  is dominated by strong CO emission bands. This complicates the correction of the telluric absorption lines. A global fit of the absorption lines in the entire K-band spectra resulted in badly over-subtracted spectra due to the strong CO emission bands at the red end of the spectrum. 
 
Instead we first fitted the spectra bluewards of 2.29 \micron\ to obtain a realistic correction for all elements except CO (which is the dominant absorber redwards of 2.29 \micron). After that we fixed the abundances of those elements as well as the wavelength shift and fitted the full spectrum allowing for only CO to vary. This resulted in a proper correction of the entire observed $K$-band spectrum. 

As this study focusses on the OB star population, we selected all the spectra with a S/N $>$ 30 that are consistent with an early spectral type. These spectra are dominated by \brg\ absorption as well as absorption lines of \hei\ and \heii\ \citep{Hanson96,Hanson05}. We did not find any stars showing photospheric \brg\ emission as seen in e.g. W49 \citep{Wu16}  in our spectroscopic dataset.
 We also added the CO emitting object to the sample as this is a massive YSO \citep{Brgspec06}.  A total sample of 15 OB-type stars and one object showing CO bandhead emission at 2.3 \micron\  was identified. Their parameters and properties will be discussed in the following sections.

\begin{figure*}[!t]
   \includegraphics[width=\hsize]{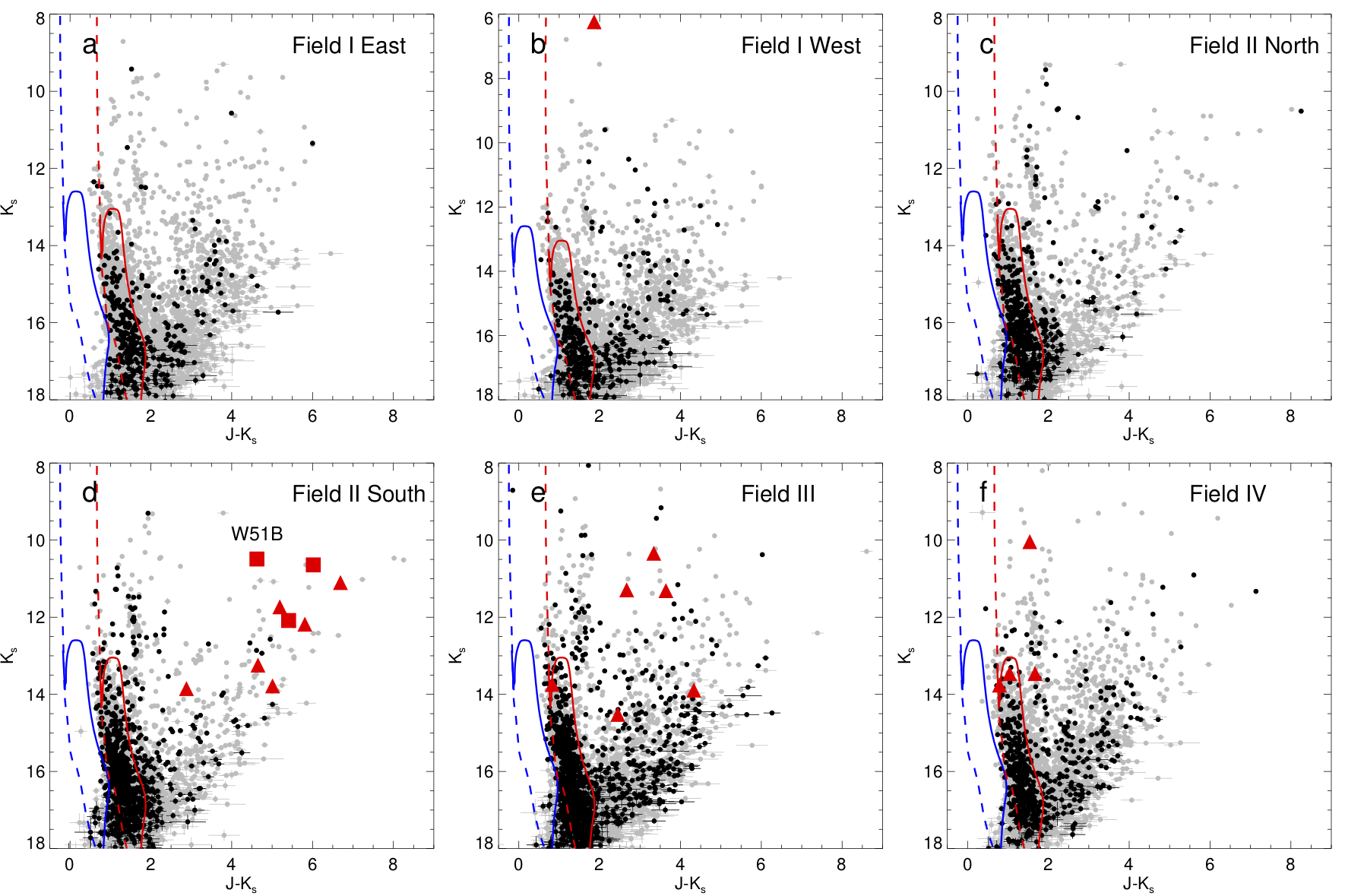}
      \caption{$J$-\Ks\ vs. \Ks\ colour-magnitude diagram of the  four observed fields in W51 showing the positions of all the stars in each field (grey points) and the sub-clusters (black points) defined by the circles on Fig \ref{fig:W51subregion}. The red triangles indicate the spectroscopically classified massive stars.  The red triangle in panel b is the position of LS1 (note the different y-axis in this panel). The red squares in panel d are the massive stars classified by \citet{Figueredo08} and \citet{Barbosa08}. The blue dashed vertical line represents the un-reddened isochrone for main sequence stars with an age of 1\,Myr \citep{Ekstrom12,Yusof13}. The blue solid line is the reddening-free  pre-main-sequence isochrone of \citet{Tognelli11}. The red lines represent the same isochrones, but reddened with \ak = 0.45 mag to highlight the foreground extinction. }
    \label{fig:cmd}
\end{figure*}

The spectra dominated by CO absorption lines, originating from late type stars (F or later), were removed from the sample as they are not produced  by massive stars belonging to the W51 region (Tab. \ref{tab:LTstars}).  Intermediate-mass pre-main-sequence stars also have CO absorption in their spectra, and do belong to the cluster \citep[e.g.][]{Bik10}, however, they are spectroscopically indistinguishable from late type foreground dwarfs or background giants. Only if their cluster membership can be assured by other means, such as  X-ray emission  \citep[e.g.][]{Feigelson13},  these objects could be confirmed as pre-main-sequence stars and included in the W51 sample.

\begin{figure*}
   \includegraphics[width=\hsize]{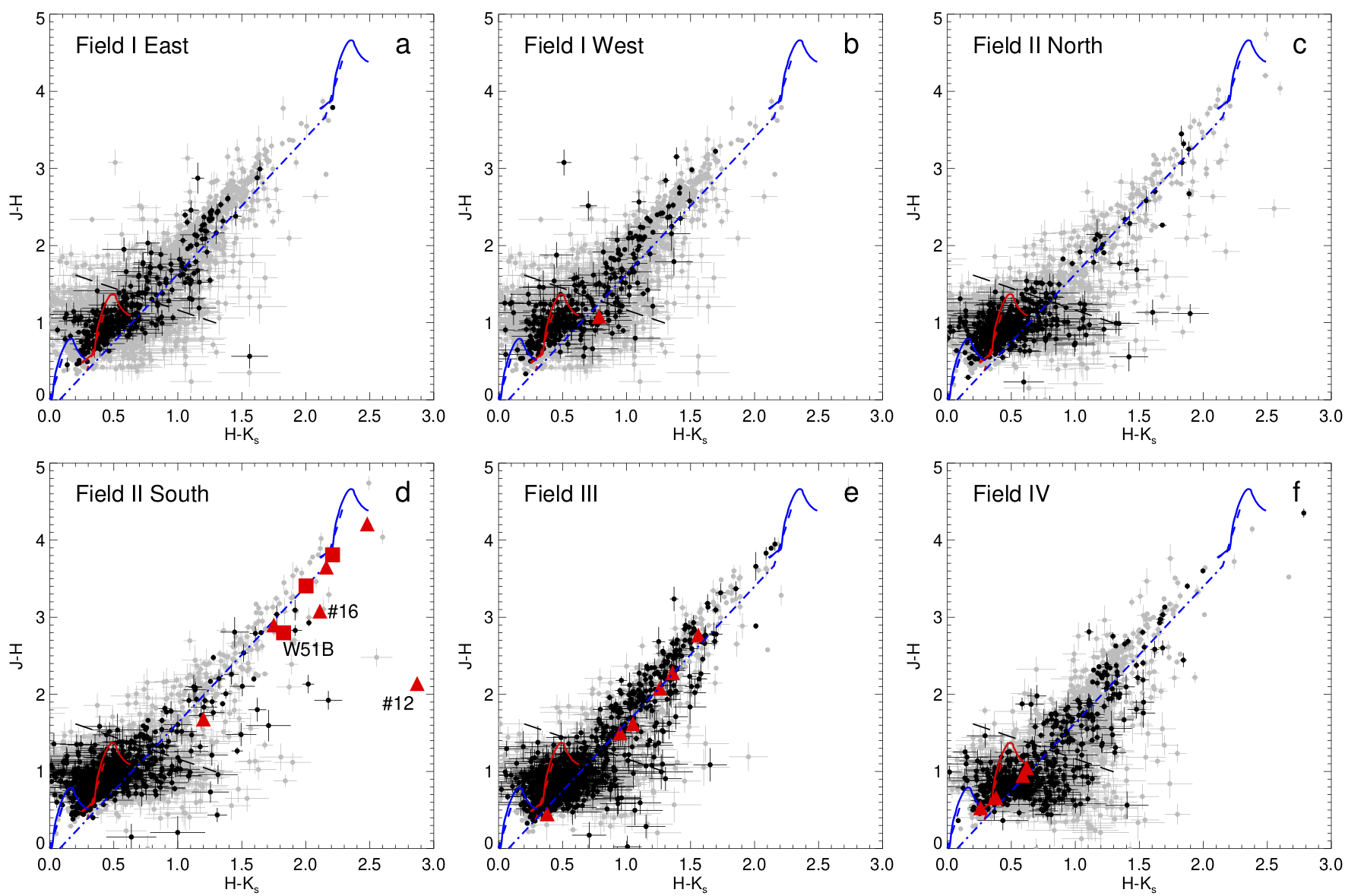}
      \caption{$H$-\Ks\ vs. $J$-$H$ colour-colour diagram of the  four observed fields in W51 showing the positions of all the stars in each field (grey points) and the sub-clusters (black points) defined by the circles in Fig \ref{fig:W51subregion}.  The red triangles indicate the spectroscopically classified massive stars. The red triangle in panel b is the position of LS1. The red squares in panel d are the massive stars classified by \citet{Figueredo08} and \citet{Barbosa08}. The blue dashed lines are the unreddened 1 Myr isochrone (around $H$-\Ks = 0) and reddened with \ak=3.0 mag (around $H$-\Ks = 2.5 mag). The blue solid lines are the unreddened 1 Myr PMS isochrones and the reddened isochrone with \ak=3.0 mag.  The blue dashed-dotted is the reddening line connecting the high-mass end of the two isochrones. The red lines represent the lines at \ak=0.45 mag.}
\label{fig:ccd}
\end{figure*}

\section{Results}\label{sec:results}
\subsection{Photometric properties}\label{sec:photom}

In this paper we focus on the analysis of the four fields for which we  obtained $K$-band spectroscopy (Fig. \ref{fig:largescale}). Two of the fields, field I and II, are located in W51A  and cover the \hii\ regions G49.58-0.038 and G49.5-0.4 respectively (Tab. \ref{tab:obslog}). Field III and IV contain two HII regions in the W51B cloud: G49.2-0.3 (field III) and G48.9-0.5 (field IV). Fields II, III and IV are also studied by \citet{Kumar04}, while the fields I and II have been studied by \citet{Okumura00} and are part of their subgroups 1, 2 and 3. 

Exploring the larger scale picture of the W51 GMC in different tracers (Fig. \ref{fig:largescale}), it becomes evident that the GMC has been fragmenting at different locations forming several star clusters 10s of parsecs away from each other. The BOLOCAM \citep{Aguirre11} 1.1 mm dust continuum (green contours in Fig. \ref{fig:largescale}) highlights the dense clumps in the GMC, while the 21 cm radio map from NVSS \citep[white contours,][]{Condon98} shows  the \hii\ regions in the GMC. Most of the \hii\ regions are associated with a dense mm core, suggesting that they harbour young, embedded clusters which have not yet fully evacuated the dust and gas they formed in. 

To get individual measurements of each cluster and their properties, we focus on  the clusters themselves. Fig. \ref{fig:W51subregion} shows zoom-ins of the clusters in our near-infrared data obtained with SOFI, which are highlighted by white circles. 
In field I, there is only little diffuse emission present. The brightest star in this field is LS1 \citep{Okumura00}, identified as a P-Cygni star \citep{Clark09}. This evolved massive star is surrounded by a diffuse shell, possibly created by its stellar wind. 
 Field II contains the youngest region G49.5-0.4. Especially the southern region in this field shows very high extinction by dust and hosts  two proto-cluster candidates \citep[W51 Main and IRS2,][]{Ginsburg12}. The cluster in field III seems to be emerging from the edge of a dense clump in the molecular cloud, while in field IV, we find two clusters inside the same HII region still surrounded by  molecular gas.

\begin{figure*}[!t]
   \includegraphics[width=\hsize]{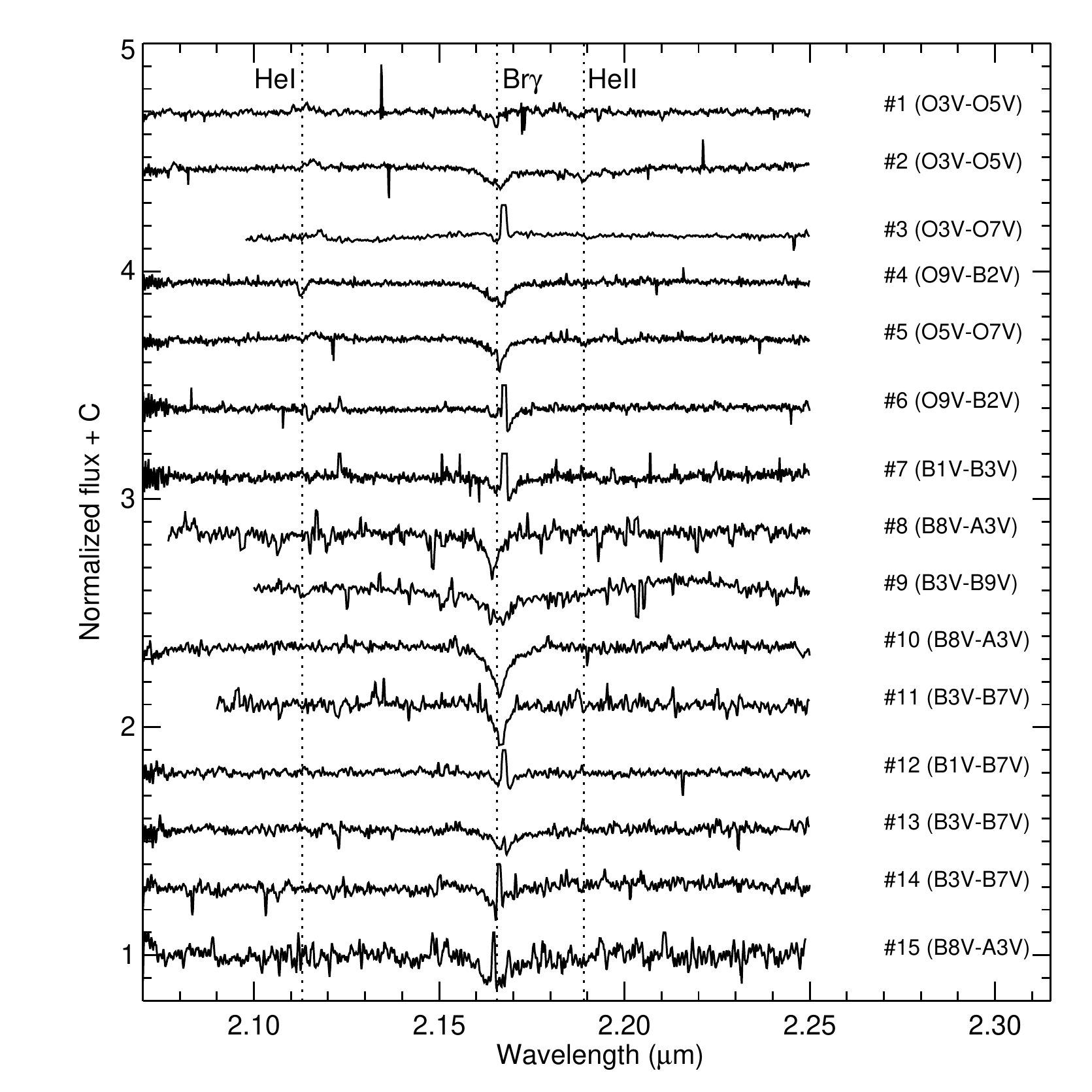}
      \caption{Normalized $K$-band spectra of OB-type stars as taken with the multi-object mode of LUCI. The star number and their derived  spectral type are indicated. Indicated with dashed lines are the diagnostic lines for the spectral classification. The narrow emission lines at 2.166 \micron\ are of nebular origin (see text). Their profiles are truncated for clarity.}
    \label{fig:OBspectra}
\end{figure*}


The white circles drawn in Fig \ref{fig:W51subregion} highlight the sub-clusters we have selected for further analysis. The size of the white circles are chosen to cover the visual extend of the clusters and their surrounding \hii\ region.
Figure \ref{fig:cmd} shows the $J$ - \Ks\ vs \Ks\ color-magnitude diagrams (CMDs) of the different clusters. In grey scale, all the stars in the  4.9\arcmin $\times$ 4.9\arcmin\ FOV of the specific field are shown and the black points are the stars inside the drawn circles on Fig. \ref{fig:W51subregion}. All sources with a detection in $JH$\Ks\ and a photometric error in all three bands less than 0.2 mag are plotted in the diagrams. The dashed line indicates the main sequence isochrone from the Geneva stellar evolution models without rotation for an age of 1 Myr \citep{Ekstrom12,Yusof13} and assuming a distance of 5.4 kpc \citep{Sato10}. The  1 Myr PMS isochrone of \citet{Tognelli11} is overplotted as a solid line.

 A mildly reddened, blue main sequence between  $J$-\Ks\ = 1 and 2 mag. is visible towards all of the clusters. This is very likely the unreddened (\ak $\lesssim$ 0.45 mag)  foreground population towards W51.  Additionally, each panel shows a reddened cluster population starting at  $J$-\Ks\ $\approx$ 2 mag and extending to $J$-\Ks\  $\approx$ 7 mag for the most reddened stars.

Figure \ref{fig:ccd} shows the $H$-\Ks, $J$-$H$ colour-colour diagrams (CCDs) of the same subfields.  The blue dashed and solid lines represent the 1 Myr isochrone as shown in the CMD unreddened around $J$-$H$ = 0 and reddened with \ak = 3.0 mag using the \citet{Nishiyama09} extinction  law around $J$-$H$ = 4. The reddening line, connecting the two isochrones is plotted as a blue dashed-dotted line. 

As in the CMD, the  foreground population is clearly detected at low extinction values as a cloud of points at low $J$-$H$ and $H$-\Ks\ values. The cluster population is visible at higher extinction, reaching values up to \ak\ $\approx$ 3.0 mag.   Apart from the spectroscopically identified OB stars it is hard to totally exclude background stars in our sample. The background stars will most likely be late type giant and AGB stars. However, by including Spitzer 8 \micron\ observation, \citet{Nandakumar18} showed that they can be distinguished from young and early type stars. As we are not attempting a full photometric analysis of the cluster population we do not attempt to remove the background stars, but assume the contribution is small due to the large extinction in W51.

Especially in field II, many infrared excess sources are visible as sources below the reddening line. This suggests that, together with the high extinction measured towards this region, this field contains  the youngest region in our sample of clusters. 
A comparison between the grey and the black points shows that there are also many sources outside the white circles with a infrared excess. These are also likely member of W51, as we drew the circles on the visual extend of the  clusters and \hii\ regions, there are very likely young stars located outside the circles. Most likely the stellar population of W51 and its sub-clusters cover all the observed frames.

\begin{table*}
\caption{Spectroscopically classified  OB stars in W51}
\label{tab:OB_classification}
\centering
\begin{tabular}{r l c c c l l l l}
\hline\hline
\# &ID & Field& Ra & Dec & J & H & K   & Sp Type\tablefootmark{b}\\
              &  &     &  (h m s)     &     (\degr\ \arcmin\ \arcsec)       & (mag) & (mag) & (mag) & \\
\hline
1&OBS7slit08 & IV&19 22 14.39  &+14 03 09.0      &11.58  $\pm$ 0.01      &10.63  $\pm$ 0.01    & 10.04 $\pm$  0.01& O3V - O5V\\
2&OBS6slit07 & III&19 23 01.68  &+14 16 40.4     & 13.68  $\pm$  0.06      &11.60  $\pm$  0.04      &10.34  $\pm$  0.04& O3V - O5V\\
3\tablefootmark{a}&OBS2slit17 & II&19 23 42.85 & +14 30 27.7   &   17.79 $\pm$   0.03 &     13.58 $\pm$   0.02      &11.10   $\pm$ 0.02  & O3V - O8V\\
4&OBS9slit17 & III&19 23 07.76  &+14 15 48.6      &13.96 $\pm$   0.02      &12.34  $\pm$  0.01     & 11.29 $\pm$   0.01&O9V - B2V\\
5&OBS9slit08 & III&19 23 02.65  &+14 16 32.9      &14.95 $\pm$   0.02      &12.67  $\pm$ 0.01     & 11.31  $\pm$  0.01&O5V - O7V\\
6&OBS3slit16 & II&19 23 41.03  &+14 29 26.9    &  17.99 $\pm$   0.05   &   14.34  $\pm$  0.01      &12.18  $\pm$ 0.01& O9V - B2V\\
7&OBS3slit17 & II&19 23 44.79  &+14 29 11.0     & 17.89  $\pm$  0.04     & 14.99  $\pm$  0.02      &13.24  $\pm$  0.02 & B1V - B3V\\
8&OBS7slit24 & IV&19 22 14.74  &+14 04 43.3      &14.50  $\pm$  0.01      &13.84 $\pm$  0.01     & 13.46  $\pm$  0.01& B8V - A3V\\
9&OBS8slit24 & IV&19 22 18.02  &+14 05 10.3      &15.13  $\pm$  0.01      &14.08 $\pm$   0.02      &13.46  $\pm$  0.02&B3V - B9V\\
10&OBS9slit23 &III &19 22 59.25  &+14 17 22.9      &14.57  $\pm$  0.01     &14.12   $\pm$ 0.02   &  13.74   $\pm$ 0.02& B8V - A3V\\
11&OBS8slit23 & IV&19 22 17.45  &+14 04 56.3      &14.55 $\pm$  0.01      &14.02   $\pm$ 0.01     & 13.76   $\pm$ 0.02& B3V - B7V\\
12\tablefootmark{c}&OBS3slit15 &II &19 23 42.99 &	+14 30 17.3 & 18.79 $\pm$ 0.08&	16.65 $\pm$ 0.02 & 13.78 $\pm$ 0.01 & B1V - B7V\\
13&OBS3slit09 &II &19 23 37.36  &+14 31 17.0    &  16.73 $\pm$   0.02  &    15.05 $\pm$   0.02      &13.85  $\pm$  0.03& B3V - B7V \\
14&OBS9slit15 &III &19 23 04.51  &+14 15 33.5      &18.22 $\pm$   0.05      &15.45  $\pm$  0.01      &13.89   $\pm$ 0.01&B3V - B7V\\
15&OBS6slit16 &III &19 23 07.49  &+14 16 05.9      &16.97  $\pm$  0.02      &15.47 $\pm$   0.02     & 14.52  $\pm$  0.02& B8V - A3V\\
\hline
16&OBS1slit17 & II&19 23 38.50 & +14 32 02.4   &   16.92 $\pm$  0.02 &     13.84  $\pm$ 0.01     & 11.73 $\pm$  0.01 & YSO\\
\hline
LS1\tablefootmark{d}  & ---&I & 19 23 47.65& +14 36 39.1 & 8.10 $\pm$ 0.02 & 7.02 $\pm$ 0.06 & \phantom{0}6.24 $\pm$ 0.02 & P Cygni SG\\
50\tablefootmark{e} & --- &II & 19 23 42.88 & +14 30 12.8 & 17.49      $\pm$   0.03          &14.09      $\pm$0.01        &  12.09       $\pm$ 0.01 & O6.5V   \\
57\tablefootmark{e} & --- &II &19 23 33.74 & +14 29 53.8  & 16.66   $\pm$         0.03    &      12.85   $\pm$        0.01   &       10.64    $\pm$      0.01  &O4 V\\
61\tablefootmark{e} & --- & II& 19 23 47.21& +14 29 43.6 & --- & 15.16	 $\pm$    0.02	&       12.51 $\pm$     0.02 & O7.5V\\
W51d\tablefootmark{e} & --- &II &19 23 39.92 & +14 31 08.5 &  15.11   $\pm$      0.02     &     12.32  $\pm$       0.02   &         10.50  $\pm$       0.03 & O3V - O4V \\
IRS2E\tablefootmark{e,f} & --- &II & 19 23 40.12 & +14 31 05.9 & --- & 16.70 $\pm$ 0.01 & 10.80 $\pm$ 0.01 & YSO\\ 
\hline
\end{tabular}
\tablefoot{
\tablefoottext{a}{Source \#44 in \citet{Figueredo08}}
\tablefoottext{b}{Spectral types obtained by comparison with reference stars from \citet{Hanson96,Hanson05} and \citet{Ostarspec05}.} 
\tablefoottext{c}{Not detected in the $J$-band in our catalogue, hence we used the value from the UKIDDS Galactic Plane Survey DR6 \citep{Lucas08}}
\tablefoottext{d}{Identified by \citet{Okumura00} and spectrally modelled by \citet{Clark09}. Photometry taken from 2MASS.}
\tablefoottext{e}{Sources spectrally classified by \citet{Figueredo08} and \citet{Barbosa08}. Photometry taken from our own data.}
\tablefoottext{f}{Photometry taken from \citet{Figueredo08}.}

}
\end{table*}

The photometric properties of the spectroscopically studied stars are listed in Tab. \ref{tab:OB_classification} and  marked in the CMD and CCD. Additionally we add the stars of which a spectroscopic classification is done in the literature: LS1 \citep{Clark09} and 4 OB stars \citep{Figueredo08,Barbosa08} to our analysis.  For LS1 we used the 2MASS photometry as the source is saturated in our images.  Source \#12 is not detected in our $J$-band images, and therefore is not listed in our catalogue. For source \#12 we use $JH$\Ks\ magnitudes  from the UKIDDS Galactic Plane Survey DR6 \citep{Lucas08}

For the OB stars from \citet{Figueredo08} and \citet{Barbosa08} we matched the sources with our photometric catalogue and used our photometry. For all but one source our photometry is consistent with that of  \citet{Figueredo08}. For W51d, however, our photometry is very different. Our $J$ and \Ks\ magnitudes are about a magnitude brighter than what \citet{Figueredo08} found, while our $H$ magnitude is even 2 magnitudes brighter. This results in a totally different $J$-$H$ and $H$-\Ks\ colour as well. \citet{Figueredo08} found a large K-band excess for this source, suggesting it has a circumstellar envelope. With our photometry that \Ks-band excess disappears (Fig. \ref{fig:ccd}), and the source ends up close to the reddening line. The reason of the difference is not clear to us, but could be related to the very crowded and high background region this source is located in. The $K$-band spectrum of this source by \citet{Barbosa08} shows a photospheric spectrum of an early O stars with \heii\ in absorption. This would be incompatible with the spectrum of a star with a strong \Ks\ excess as the faint \heii\ absorption line would be invisible due to veiling by the circumstellar material. Therefore, we chose to use our photometry  for  the analysis of this star.  The YSO IRS2E,  whose spectrum is presented in \citet{Barbosa08}, is only detected in $H$ and \Ks. As our catalogues only have detections in $JH$\Ks\ we took its $H$ and \Ks\ magnitudes from \citet{Figueredo08}.

In Figure \ref{fig:cmd}, panel b, we show Field I where LS1 is indicated with a red triangle, however, we do no detect any other massive stars. Most spectroscopically identified OB stars are located in the southern cluster of field II. This is the location of the most active star forming complex in W51: G49.5-0.4.  The OB stars from \citet{Figueredo08} are shown as red squares. The location of the OB stars in the CCD (Fig. \ref{fig:ccd}) shows that in field II all spectroscopically identified OB stars are highly reddened.  In the two \hii\ regions in W51B (fields III and IV) we also identify several OB stars in each region.  Also in field III most  OB stars in our spectroscopic sample are reddened (apart from 1 object), while in field IV, all OB stars have very low reddening. 
In the CCD (Fig. \ref{fig:ccd}), we have two spectroscopic targets which shows a significant infrared excess. Star \#16 is located in field II, south and has a $H$-\Ks\ colour of 2.1 mag.  With $J$-$H$ = 3.08 mag this results in a mild infrared excess. The source with an  extreme infrared excess is source \#12.

\begin{figure*}[!ht]

   \includegraphics[width=\hsize]{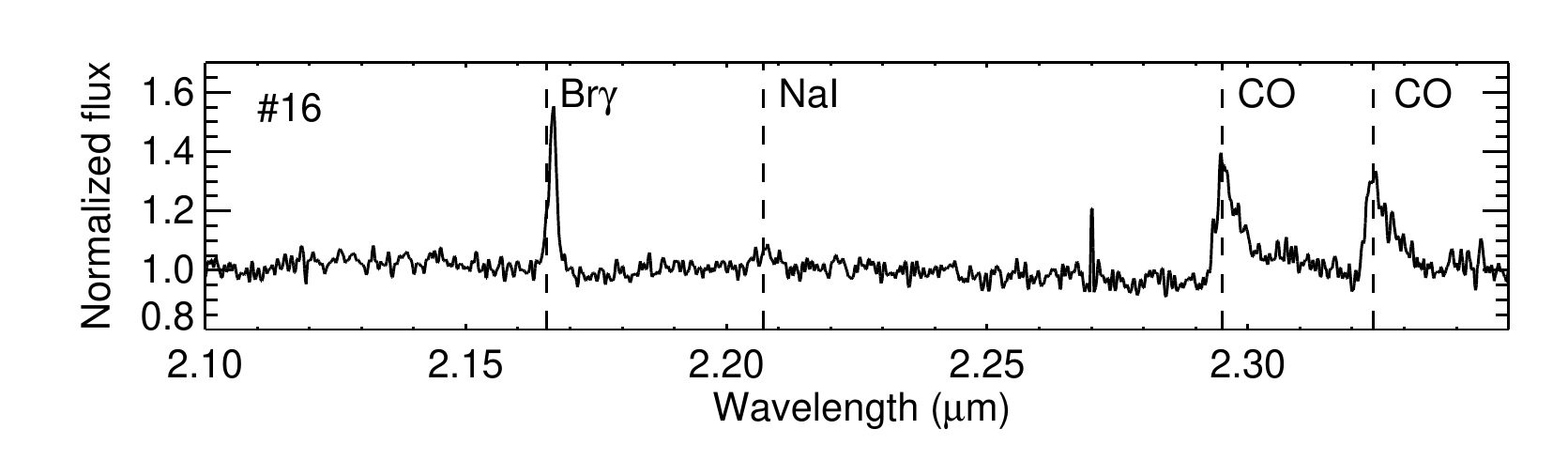}
      \caption{Normalized $K$-band spectrum of source \#16, identified as YSO based on the CO bandhead emission. The dashed lines indicate the emission lines in the spectra.}
    \label{fig:YSO}
\end{figure*}

\subsection{Spectral Classification}

From our total spectral sample we identified 15 stars with early spectral type $K$-band spectra (Tab. \ref{tab:OB_classification}). Their normalized $K$-band spectra are shown in Fig.\,\ref{fig:OBspectra}. All the $K$-band spectra are dominated by broad \brg\ in absorption. The narrow emission lines visible in several stars are due to  nebular emission from the \hii\ region. We tried to correct for the nebular contamination by fitting and subtracting the background around the positions of the spectra, but residual of the nebular emission are still present in several spectra. The spectra are not corrected for the systemic velocity of the earth, therefore some of the lines might appear shifted. 

For the earlier type stars \hei\ (2.113 \micron) and \heii\ (2.189 \micron) absorption lines are present as well as a N{\sc iii} emission complex at  2.115 \micron\ \citep{Hanson96,Hanson05}. To determine the spectral types of each star, their spectra are visually compared to high resolution $K$-band spectra of reference OB-type stars with optical classification from \citet{Hanson05} and \citet{Ostarspec05}, while for spectral types mid-B and later the low-resolution spectra of \citet{Hanson96} are used. 

We performed spectral classification similar to \citet{Bik12} as follows; high resolution, high S/N reference spectra are degraded in resolution and S/N to match the observed LUCI spectra of W51. After that a visual comparison is made between the reference spectra and the observed spectra to identify the range of best matching reference spectra and therefore best matching spectral types. The resulting spectral types are given in Tab. \ref{tab:OB_classification} and Fig. \ref{fig:OBspectra}.

From the total sample of 15 OB stars, we identified six O stars. Three stars, \#,2 and 3, are identified as early O stars, based on the lack of \hei\ absorption and the presence of \niii\ emission. Star \#3 shows a strong contamination of \brg\ emission from the \hii\ region. This also makes the use of the \hei\ line for the classification uncertain, as this line could be contaminated by nebular emission. We do not see any \hei\ absorption in the spectra, suggesting a spectral type of O5V or earlier, however, the \heii\ line is relatively weak, which suggests a spectral type as late as O8V. For this reason we classified this star as O3V - O8V. Star \#3 is also included in \citet{Figueredo08} as star \#44 in their paper, classified as an O5V star, consistent with our classification.

The remaining nine stars are classified as B stars.  Their spectral classification  relies on the \hei\ and \brg\ lines, while for the late-B stars only the strength of the \brg\ absorption is used as no other lines are present.   Most stars show a nebular emission line on top of the \brg\ absorption line. This affects the accuracy of the spectral classification, especially for the B stars where \brg\ is the most important line, as we can only use the absorption wings to estimate the strength. This results in a less accurate classification which is reflected in the large uncertainty in the spectral classification. 


The spectrum of star \#16 shows an emission line spectrum instead of a photospheric absorption line spectrum  (Fig. \ref{fig:YSO}). This object shows strong \brg\ emission as well as CO first-overtone emission redwards of 2.29 \micron. The CO emission is seen in several (massive) YSOs  and young stars with remnant accretion disks in many  star forming regions \citep[e.g.][]{Chandler95,COletter04,Blum04,Brgspec06,Bik12,Puga10,Wheelwright10,Stolte10,Ellerbroek13,Ilee13,Cooper13,Pomohaci17}. Also the K-band spectrum of IRS2E from \citet{Barbosa08} has a similar spectrum to that of our source \#16.
 The CO emission originates in neutral gas with a temperature of $\sim$2000K located in the inner disk \citep{COletter04,Wheelwright10}.
 The \brg\ emission line most likely originates from several different environments in the circumstellar matter such as the ionized surface of the disk, disk wind or the surrounding \hii\ region \citep{Brgspec06}.  Additionally, faint \ion{Na}{i} is detected at $\sim$2.208 \micron\ consistent with warm neutral gas. This emission is detected in other YSOs as well \citep{Cooper13}.  

Star \#12 shows a strong near-infrared excess in the photometry, however, the $K$-band spectrum shows an absorption line in \brg. This lead to the classification of B1V - B7V. The strong infrared excess from the photometry is not compatible with the $K$-band spectrum. Object \#16 only shows a small  excess in the photometry and already shows a $K$-band spectrum dominated by the circumstellar matter.  We therefore exclude star \#12 from the analysis of the HRD as the magnitudes might be affected by an infrared excess, which would result in wrong position in the HRD.

The very luminous source LS1 is also covered by our spectroscopic survey, however, the $K$-band spectrum is partly saturated. Spectral modelling by \citet{Clark09} of their K-band spectrum shows that LS1 is a P Cygni type supergiant with an expanding envelope \citep[see also][]{Okumura00}. 
\citet{Clark09} derived physical parameters for 3 different distances for LS1; 2, 3.4 and 6 kpc. They excluded the 2 kpc distance due to the fact that the derived mass would be too low to be a P Cygni supergiant.  Due to the association with W51, the 6 kpc distance is favoured, bringing the mass in the more typical P Cygni supergiant regime. In the remaining of the paper we will use the parameters derived for 3.4 kpc and 6 kpc (Tab. \ref{tab:LS1}). The four O stars found by \citet{Figueredo08} are all  early type O stars, significantly increasing the number of spectroscopically identified O stars in W51.

\begin{table}[!t]
\caption{Physical properties LS1}
\label{tab:LS1}
\centering
\begin{tabular}{l l | r l}
\hline\hline
Distance: & 6 kpc & Distance: &3.4 kpc\\
log(\teff): &  4.121 & log(\teff):  &4.127  \\
log(L/L$_{\sun}$) & 5.75 & log(L/L$_{\sun}$): & 5.30\\
\ak & 1.2 mag & \ak & 1.2 mag \\
Mass: & $\sim$25 M$_{\sun}$ & Mass: & $\sim$40 M$_{\sun}$ \\
\hline
\end{tabular}
\tablefoot{Paremeters taken from \citet{Clark09}.}
\end{table}

\begin{table*}[!h]
\caption{Physical parameters of the OB stars in W51}
\label{W51startable}
\centering
\begin{tabular}{r c l l l ll l l l}
\hline\hline
\# & Field &  log($T_{\mathrm{eff}}$) & \ak(N09)\tablefootmark{a} &\ak(I05)\tablefootmark{b}  & log($L$)(N09)\tablefootmark{a} & log($L$)(I05)\tablefootmark{b} &  Mass\tablefootmark{a} &  Mass\tablefootmark{b} & Avarage Mass \\
  & & (K)  &   (mag)       & (mag)   &  (\lsun) & (\lsun)& (\msun) & (\msun)& (\msun)\\
\hline


  1  & IV&   4.63   $\pm$   0.02 &  0.94   $\pm$  0.05    & 1.25   $\pm$   0.18     & 5.62   $\pm$    0.08       &   5.74    $\pm$    0.10  & \phantom{0}50$^{+5}_{-2}$ & \phantom{0}58$^{+8}_{-7}$ &\phantom{0}54$^{+9}_{-7}$ \\
  2   & III&  4.63   $\pm$   0.02 &  1.85   $\pm$   0.12   & 2.47   $\pm$   0.37     & 5.86   $\pm$   0.09        &   6.11    $\pm$    0.17  & \phantom{0}68$^{+8}_{-8}$ & \phantom{0}95$^{+25}_{-20}$&\phantom{0}77$^{+26}_{-22}$\\
3   &II &    4.60  $\pm$    0.05 &  3.54  $\pm$   0.17    & 4.70  $\pm$   0.68      &  6.13  $\pm$    0.17       &   6.59   $\pm$    0.32   & \phantom{0}95$^{+30}_{-20}$ & 200$^{+100}_{-80}$ &148$^{+105}_{-82}$ \\\
4   &III&    4.43  $\pm$    0.10 &  1.53  $\pm$  0.09      & 2.03  $\pm$    0.30    & 4.77  $\pm$     0.21       &    4.97   $\pm$     0.24 & \phantom{0}17$^{+3}_{-2}$ & \phantom{0}20$^{+5}_{-3}$ &\phantom{0}19$^{+6}_{-4}$\\
5   &III&    4.59  $\pm$    0.02 &  1.99  $\pm$  0.10      & 2.65  $\pm$    0.39    & 5.41  $\pm$    0.09        &    5.67   $\pm$     0.17 & \phantom{0}38$^{+3}_{-3}$ & \phantom{0}50$^{+12}_{-9}$ &\phantom{0}44$^{+12}_{-9}$\\
6    &II&   4.43  $\pm$    0.10 &  3.04  $\pm$   0.15     & 4.04  $\pm$    0.59    & 5.02  $\pm$     0.22       &    5.42   $\pm$     0.32 & \phantom{0}20$^{+5}_{-2}$ & \phantom{0}30$^{+18}_{-7}$&\phantom{0}25$^{+18}_{-7}$\\
7     &II&  4.33  $\pm$    0.09 &  2.45  $\pm$   0.12     & 3.25  $\pm$    0.48    & 4.16  $\pm$     0.23       &    4.48   $\pm$     0.29 & \phantom{0}11$^{+2}_{-2}$ & \phantom{0}13$^{+3}_{-2}$&\phantom{0}12$^{+4}_{-3}$\\
8    &IV&   4.02  $\pm$    0.08 &  0.50  $\pm$  0.04      & 0.67  $\pm$    0.11    & 2.36  $\pm$     0.18       &    2.43   $\pm$     0.19 & \phantom{00}3.5$^{+0.5}_{-0.5}$ & \phantom{00}4$^{+0.5}_{-0.5}$ & \phantom{00}3.8$^{+0.7}_{-0.7}$\\
9    & IV  &4.14  $\pm$    0.10 &  0.86  $\pm$  0.06      & 1.15  $\pm$    0.18    & 2.85  $\pm$     0.25       &    2.97   $\pm$     0.25 &  \phantom{00}4.5$^{+0.5}_{-0.5}$ & \phantom{00}5$^{+0.5}_{-0.5}$& \phantom{00}4.8$^{+0.7}_{-0.7}$\\\
10   &III &  4.02  $\pm$    0.08&  0.49  $\pm$  0.04      &  0.65  $\pm$    0.11    &  2.24  $\pm$     0.18       &  2.30   $\pm$     0.19  & ---  \tablefootmark{c} & --- \tablefootmark{c}&  --- \tablefootmark{c}\\
11   & IV & 4.19   $\pm$   0.04&  0.36   $\pm$  0.03     &  0.48   $\pm$   0.08    &  2.70   $\pm$    0.12       &  2.75    $\pm$    0.12  & \phantom{00}4.5$^{+0.3}_{-0.3}$ & \phantom{00}4.8$^{+0.3}_{-0.3}$ & \phantom{00}4.7$^{+0.4}_{-0.4}$\\
12   &II  & 4.30   $\pm$    0.10&  3.96   $\pm$   0.19    &  5.27   $\pm$   0.77    &  4.41   $\pm$    0.34       &  4.93    $\pm$    0.45 &  \phantom{0}13$^{+3}_{-3.5}$ & \phantom{0}17$^{+11}_{-4}$  &\phantom{0}15$^{+11}_{-5}$  \\
13   & II & 4.19   $\pm$   0.04&  1.66   $\pm$ 0.09      &  2.20   $\pm$   0.33    &  3.18   $\pm$    0.12       &  3.40    $\pm$    0.18  & \phantom{00}5.5$^{+0.5}_{-0.5}$ & \phantom{00}6.5$^{+0.5}_{-0.5}$ & \phantom{00}6.0$^{+0.7}_{-0.7}$\\
14   & III & 4.19   $\pm$   0.04&  2.16   $\pm$  0.10     &  2.87   $\pm$   0.42    &  3.37   $\pm$    0.12       &  3.65    $\pm$    0.20  & \phantom{00}6.2$^{+0.5}_{-0.5}$  & \phantom{00}7.5 $^{+1}_{-0.7}$& \phantom{00}6.9 $^{+1.1}_{-0.9}$\\
15   & III & 4.02   $\pm$   0.08&  1.28   $\pm$ 0.07      &  1.69   $\pm$   0.25    &  2.24   $\pm$    0.18       &  2.41    $\pm$    0.21  & \phantom{00}3.0$^{+0.5}_{-0.2}$ & \phantom{00}3.5$^{+0.5}_{-0.5}$ & \phantom{00}3.3$^{+0.7}_{-0.5}$ \\
\hline
50\tablefootmark{d} & II&	4.58   $\pm$    ---&  2.88 $\pm$ 0.14 &3.82 $\pm$ 0.56 & 5.42 $\pm$    --- & 5.80 $\pm$    --- & \phantom{0}37 & \phantom{0}60 & \phantom{0}49\\
57\tablefootmark{d}  &II& 4.63  $\pm$    ---& 3.16 $\pm$ 0.15 & 4.20 $\pm$ 0.61 & 6.26 $\pm$    ---& 6.68 $\pm$    --- & 120 & 200 & 160\\
61\tablefootmark{d}  &II& 4.58  $\pm$    ---& 3.76 $\pm$ 0.18 & 5.00	$\pm$ 0.73  & 5.60 $\pm$  ---  & 6.10 $\pm$  ---  & \phantom{0}45 & \phantom{0}90 &\phantom{0}68\\
W51d\tablefootmark{e}&II & 4.63  $\pm$    ---&2.63 $\pm$ 0.14 & 3.50 $\pm$ 0.52 & 6.11 $\pm$    ---& 6.46 $\pm$    --- & 100 & 170 & 135 \\

\hline\label{tab:logl_logt}
\end{tabular}
\tablefoot{
\tablefoottext{a}{Stellar properties derived assuming the extinction law of \citet{Nishiyama09}}
\tablefoottext{b}{Stellar properties derived assuming the extinction law of \citet{Indebetouw05}} 
\tablefoottext{c}{Possible foreground object (see further discussion in the text).}
\tablefoottext{d}{Spectral types and photometry taken from \citet{Figueredo08}. There is no uncertainty quoted on the spectral type, therefore no errors are given for the $T_{\mathrm{eff}}$ and $L$.}
\tablefoottext{e}{Spectral types and photometry taken from \citet{Barbosa08}. There is no uncertainty quoted on the spectral type,therefore no errors are given for the $T_{\mathrm{eff}}$ and $L$.}
}
\end{table*}

\subsection{Cluster membership}\label{sec:members}
Before placing the stars in the HRD in order to compare them to evolutionary tracks, possible fore- and background stars need to be removed from the sample. The W51 complex is located rather isolated in a spiral arm with very little foreground material. The CO observations show that the W51 GMC is observed in velocity space between 50 and 70 \kms\ \citep[e.g.][]{Carpenter98}, with two additional features in the CO spectra  at 7 \kms\ and between 20 and 25 \kms. These two features are associated with foreground cloud of which the 7 \kms\ cloud is a diffuse molecular cloud and the 20-25 \kms\ mostly consist of atomic neutral hydrogen detected in absorption towards IRS2 inside W51A \citep{Indriolo12}. No signs of active star formation has been reported for those clouds. All the \hii\ regions are associated with the molecular material between 50 and 70  \kms\ \citep[see for a sketch Fig. 11 of][]{Ginsburg15}.

This makes the presence of foreground O stars rather unlikely, as they are young enough to still show part of their birth material around them. It is more likely to find foreground B stars, as they live longer and are therefore not anymore associated with molecular clouds. Foreground stars would  have a much lower extinction than the cluster members as the latter are still partially embedded in the GMC. Any background source on the other hand would have a much higher extinction than the cluster members as they are behind the GMC. 

As discussed in Sect. \ref{sec:photom} we found five sources with very low extinction, possibly consistent with a less reddened foreground population. Source \#10 is the only OB stars  in field III with an \ak\ below 1 mag (\ak  = 0.49 $\pm$ 0.04 mag), while all the other OB stars have \ak\ =1.3 mag or higher. Therefore we classify source \#10 as a potential foreground star.  The other stars with low \ak\ measurements are all located in field IV (G48.9-0.3).  Here all stars have consistently \ak\ values below 1 mag., even the early O star source \#1, which is clearly located in the centre of one of the two clusters. The low extinction towards these stars could be explained by the fact that G48.9-0.3 seems to be located in front of the 68 \kms\ cloud \citep{Ginsburg15} instead of inside \citep{Kang10}. Therefore we assume that all OB stars inside G48.9-0.3 are cluster members.

\begin{figure*}[!t]
   \includegraphics[width=\hsize]{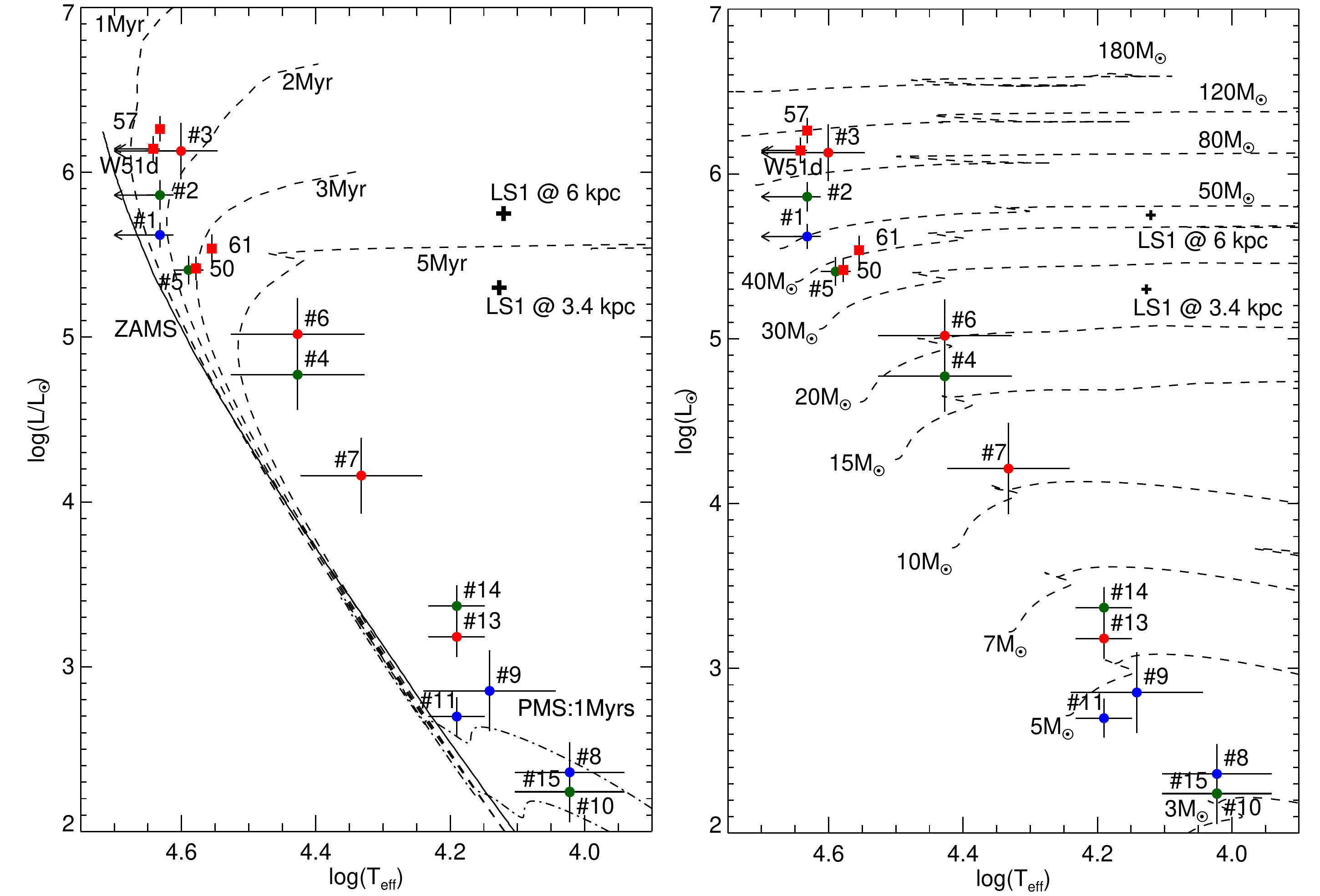}
      \caption{HRD of stars identified as OB stars in our spectroscopic survey. The coloured sources are the locations with the extinction correction of \citet{Nishiyama09}. The colour coding of the sources indicates in which region they are located: red: field II, green: field III and blue: field IV. The location of LS1 \citep{Okumura00}  for the assumed distances of 3.4 and 6 kpc is marked with a "plus" sign.   The red squares are the massive stars classified by \citet{Figueredo08} and \citet{Barbosa08}. 
      \emph{Left:} The solid line represents the ZAMS isochrone from \citet{Lejeune01} and the dashed lines the main-sequence isochrones for 1, 2, 3 and 5 Myr from \citet{Ekstrom12} and \citet{Yusof13}. The dash-dot lines represent the PMS isochrones of 1 and 2 Myr from \citet{Tognelli11}.  \emph{Right:} The dashed  lines represent the evolutionary tracks from \citet{Ekstrom12} and \citet{Yusof13}. 
      }
    \label{fig:HRD}
\end{figure*}

\subsection{HRD}

Using stellar atmosphere models we can derive a temperature and luminosity for each star  we have spectrally classified. This allows us to compare those values to isochrones and stellar evolution tracks to derive stellar masses as well as putting constraints on the star formation history of W51. We used  \citet{Martins05} to calculate the effective temperature for the O stars.  The bolometric correction and the intrinsic $H$-\Ks\ colours are taken from \citet{Martins06}. The effective temperature, bolometric correction and the intrinsic $H$-\Ks\ colours of B type main sequence stars are from \citet{Pecaut13}. The derived temperatures are listed in Tab. \ref{tab:logl_logt}. The errors in the effective temperature are determined by the uncertainties in the spectral typing.

With the intrinsic $H$-$K$ colour we were able to derive the extinction \ak\ of the stars. As shown in \citet{Bik12} and \citet{Wu14} the choice of extinction law can have a large influence on the derived luminosity and mass, especially at very high \ak. 
We tested several extinction laws in the CCD to see if the slope of a given extinction law is consistent with observed reddened main sequence.  The extinction laws of \citet{Roman07} and \citet{Cardelli89} are too shallow and the reddening lines are not following the reddened main sequence in the CCD. The extinction law of \citet{Fitzpatrick99} is too steep. The extinction laws of \citet{Rieke85} and \citet{Indebetouw05} are very similar in the near-infrared and provide a good fit to the reddened main sequence. The last extinction law we tested, \citet{Nishiyama09} also provides a good fit to the data. This last extinction law is shown in the CCD, as explained in Fig \ref{fig:ccd}.  Apart from the slope of the extinction, the ratio of the total over selective extinction ($R_{\lambda} = A_{\lambda}/E(H-K_{\mathrm{s}})$) is important when calculating the absolute extinction \ak. This cannot be constrained by $JH$\Ks\ data alone.  In Tab. \ref{tab:logl_logt} the values for \ak\ listed for both the \citet{Nishiyama09} and \citet{Indebetouw05} extinction laws are given. The \citet{Indebetouw05} law results in higher extinction values due to a higher $R_{\mathrm{K}}$  \citep[1.82 mag instead of 1.44 mag as in ][]{Nishiyama09}

We derived the remaining stellar parameters for both the \citet{Indebetouw05} and \citet{Nishiyama09} extinction laws. The absolute $K$-band magnitudes were calculated by subtracting $A_{K}$ and the distance modulus (13.66 $\pm$ 0.12 mag) from the apparent magnitudes. Considering the bolometric correction from stellar models \citep{Martins06}, we derived the bolometric magnitudes and then derived the luminosity of the OB stars (Tab \ref{tab:logl_logt}). The points dereddened using the  \citet{Indebetouw05} law all have higher luminosity due to the higher extinction.

In Fig. \ref{fig:HRD} we plot the sources in the HRD together with  the location of the zero age main sequence (ZAMS) from  \citet{Lejeune01} as well as the 1, 2, 3 and 5 Myr  isochrones (left panel) and evolutionary tracks without rotation (right panel) from  \citet{Ekstrom12} and \citet{Yusof13} as well as the pre-main-sequence isochrones from \citet{Tognelli11}. For clarity we only show the HRD with the \citet{Nishiyama09} extinction law. The HRD with de-reddening using the \citet{Indebetouw05} extinction law is presented in Fig.  \ref{fig:HRD_Ind}.    The colour coding of the points in the HRD reflects the fiels in which the sources are found. The green (field III) and blue (field IV) are originating from W51B, while the red (field II) and yellow (LS1 in field I) are located in W51A. 

The masses of the stars are calculated by interpolating the tracks to the position of the stars in the HRD (Fig. \ref{fig:HRD}, right panel). The errors on the masses are determined from the range of evolutionary tracks consistent with the position of the star in the HRD taking into account the error in luminosity and temperature.
 The resulting masses are given in Tab. \ref{tab:logl_logt} for both extinction laws. For the determination of the mass, the choice of extinction law is important and the final mass can vary a lot. For example for star \#3  the masses we derive differ by a factor 2 from each other (100 vs. 200 \msun). 
The final column of Tab. \ref{tab:logl_logt} provides the average of the masses derived for the two extinction laws.

Also the O stars from \citet{Figueredo08} and \citet{Barbosa08} are included in the HRD. We use the  spectral types and the magnitudes in Table \ref{tab:OB_classification} and calculate their luminosity, effective temperature and mass in the same way as the other stars. As for most of these stars no range in spectral types is given, we plot the position without error bar. Also LS1 is added to the HRD, where the luminosity and effective temperature are taken from Table \ref{tab:LS1}. 

For sources \#1, \#2, \#3 and W51d, the left error bar is plotted as an arrow. The reason for this is that the spectral type corresponding to that error bar is O3V  (log $T_\mathrm{eff}$  = 4.652), which is the earliest spectral type listed in the tables of \citet{Martins05} and with existing near-infrared classification criteria \citep{Hanson05}.  This means that in our HRD there are  no points to the left of log $T_\mathrm{eff}$  = 4.652, making the upper temperature error of these stars in a lower limit.  Stars hotter than log $T_\mathrm{eff}$  = 4.652 will be classified as O3V star with the current near-infrared classification. This degeneracy can be solved by spectral fitting with stellar atmosphere models.

Our sample includes stars from early type O stars, down to late B-early A stars, spanning five orders of magnitude in luminosity in the HRD. We identify eight stars of early and mid-O spectral type with masses above  30 \msun\ in the W51 GMC, of which most are located in the G49.5-0.4 \hii\ region (field II).

The sources with late O to early B spectral types (\#4, 6, 7 and 12) are located  a bit above the main sequence, however, all within 2 sigma away from the stellar isochrones.   Sources \#13 and 14 are located even further from the main sequence isochrones. Source \#15 is consistent with being a 1-2 Myr old pre-main-sequence star. We discuss this displacement in detail in Sect. \ref{sec:age}.


\begin{table}[!t]
\caption{\hii\ regions associated with massive stars}
\centering
\begin{tabular}{c l r r l l r}
\hline\hline 
 ID& \hii\ region  &Radio ID & star & radio\tablefootmark{a} & IR  \\
 &   & &  & Sptype & Sptype \\
\hline
II &G49.5-0.4& e20/e21\tablefootmark{b} & \#3	& -- & O3-O8V \\
 && c1 & \#6	& O6V & O9-B2V \\
 && e7 & \#7	& B0V & B1-B3 \\
 && b & 57	& O4V & O5 \\
&& d &W51d 	& O5V	& O3-O4 \\
III &G49.2-0.3 	&   ---  & \#2 & O4V & O3-O5V\\
IV & G48.9-0.3 & --- & \#1  & & O3-O5V\\
\hline
\end{tabular}\label{tab:radio}
\tablefoot{\tablefoottext{a}{\citet{Mehringer94}}
\tablefoottext{b}{\citet{Ginsburg16}}
}
\end{table}

\subsection{Energy budget \hii\ regions}\label{sec:radio}

The W51 GMC harbours numerous \hii\ regions, each ionized by one or more  massive stars. 
The radio free-free emission observed from these \hii\ regions can be used as a proxy for the spectral type of the ionizing sources. Most of the radio observations have been focussed on the most active site of star formation: the W51A cloud. \citet{Mehringer94} performed high resolution VLA observations with a spatial resolution of 3.1\arcsec\ at 3.6 cm of the W51A cloud and covered our fields II and III. \citet{Ginsburg12} zoomed in the central part of W51A and presented observations with a spatial resolution of 0.3\arcsec\ at 2 and 6 cm.

We compared the location of the spectroscopically identified OB stars with these radio observations and found for five stars in field II the corresponding \hii\ region (Tab. \ref{tab:radio}). Additionally, the clusters in field III and IV have their own \hii\ region and we only list their most massive stars as they are the dominant source of ionizing photons. 

When comparing the spectral types derived from the radio free-free emission with that of the stars inside the \hii\ regions, we found in most cases good agreement, thus confirming that we have identified the main ionizing sources of those regions. In two regions (G49.5-0.4 c1 and e7) we find that the radio spectral type is significantly earlier than the spectrum of the star inside the \hii\ region. This suggests that another massive stars, possibly more deeply embedded in this HII region, is responsible for its ionization.

The dust cocoons around several of these \hii\ regions are also detected in the mid-infrared imaging of \citet{Lim19}, suggesting that these \hii\ regions harbor very young objects. \hii\ region e7 around star \#7 is even identified as a massive YSO candidate. The other \hii\ regions are more extended, possible suggesting a slightly more evolved nature.

Also, there are many O stars without identified \hii\ region and vice versa. This is because of two reasons. The radio emission in this area is very complex and due to confusion as well as the fact that radio interferometers typically are sensitive to a limited range in spatial scales, it might well be that some of the fainter \hii\ regions are missed. Our spectroscopic census is not complete and several bright point sources have not been observed, therefore it is very likely that we missed some O stars.


\section{Discussion}
\label{W51:discussion}

We  obtained $K$-band spectroscopy and near-infrared imaging of members of four \hii\ regions in the W51 GMC complex. Below we discuss the implications of our findings for our understanding of the star formation history for each individual region and W51 as a whole. Table \ref{regiontable} summarizes the basic properties of the four observed clusters.

\subsection{Age of the stellar population}\label{sec:age}

We can compare the position of the massive stars in W51 in the HRD (Fig. \ref{fig:HRD}) with the location of the theoretical isochrones and derive constraints on the age of the stellar population. The early O stars, sources \#1, 2, W51d, 57 and 3 are all located between the ZAMS and the 2 Myr isochrones.  Sources \#5, 50 and 61 are a bit to the right of the 2 Myr isochrone and, especially 61,  more consistent with the 3 Myr track. 

However, several of these stars have a very high extinction and a change of extinction law changes their position in the HRD significantly. Comparing the HRD made with the \citet{Nishiyama09} extinction law (Fig. \ref{fig:HRD}) with the diagram created with the \citet{Indebetouw05} extinction law (Fig. \ref{fig:HRD_Ind}), the position of \#5, 60 and 61 all move upward  and become more consistent with the 2 Myr isochrone.

The sources with  late B to early A spectral types overlap in the HRD with the PMS tracks and can also be used to constrain the age of the stellar population. Sources \#8 and \#15 would be classified as intermediate mass pre-main-sequence stars with an age of ~ 1-2 Myrs, consistent with the age found for the  massive stars. Source \#10 was classified as foreground star due to it's very low extinction compared to the other stars in Field III. 

We have detected 3 early-mid B stars in W51 (\#4,\#6, \#7 )  as well as 2 late B stars (\#13 and \#14) whose location in the HRD (for both extinction laws) is not consistent with the above derived age of 1-3 Myrs.  Sources \#4, \#6 and \#7 are 1.5 - 2$\sigma$ away from the 1-3 Myr isochrones, while sources 13 and 14 are more than 3$\sigma$ away.
Also in other star forming regions we found that the B stars are not matching with the same isochrones as the O stars \citep{Bik12,Wu16}.  From all the 5 sources, only 1 source (\#13) is located slightly to the right of the reddening line in the CCD (Fig. \ref{fig:ccd}), suggesting that it has as small color excess ($\sim$0.25 mag in (H-K). However, this would only result in a 25 \% increase in luminosity, not explaining the large offset of source \#13 from the isochrones.

The classification of the B stars relies on just one (late B) or two (early- mid B) absorption lines \citep{Hanson96}. This is reflected in the large error bars of the stars in the HRD, however, cannot explain the large offset. For sources \#4, 6 and 7 the observed luminosity is consistent with them being late O stars, this is however not compatible with just the detection of \brg\ and \hei. 

In \citet{Bik12} several possibilities are listed causing a displacement in the HRD. It has become clear that most massive stars are observed as binaries \citep[e.g.][]{Apai07,Sana12,RamirezTannus17,Sana17}. The combined luminosity would be a factor 2 higher in the case of an equal mass binary, again only explaining part of large offset in luminosity observed. Another possibility is that the stars are not part of W51, but are located in the foreground. Sources \#6 and \#7 are associated with radio sources c1 and e7 (see Sect. \ref{sec:radio}). Additionally, the measured foreground extinction towards all the sources is similar to the other stars in W51. As shown in Sect. \ref{sec:members} there is very little molecular material along the line of sight towards W51 and non of that is associated with star formation. This makes the stars most likely member of W51.

The other possibility is that the B stars are in a different evolutionary state than the O stars. Stars \#4 and \#6 overlap within their error bars with the 5 Myrs isochrone. This would make their age consistent with LS1. Alternatively, these stars could be massive pre-main-sequence stars. Theoretical simulations of massive stars  predict the presence of massive PMS stars \citep[e.g][]{Hosokawa09}. Deep optical and near-infrared studies of high-mass star forming regions have revealed several candidates for this evolutionary phase \citep{Ochsendorf11,Bik12,RamirezTannus17}. Especially, sources \#6 and \#7 could be good candidates, as they are surrounded by a compact \hii\ region, still embedded in their natal dust cocoon, suggesting they are young objects. Follow-up optical spectra would allow a better spectral classification as well as a determination of their surface gravity. These observations would provide more constraints on their evolutionary state.

The black plus signs show the location of LS1 assuming a distance of 3.4 and 6 kpc distance, where the parameters are taken from  \citet{Clark09} and listed in Tab. \ref{tab:LS1}. The position in the HRD made \citet{Clark09} conclude that LS1 is at least 3 Myrs old, with a likely age between 3 and 6 Myrs for the furthest distance. This would suggest an age difference between LS1 and the  O type stars in W51 and therefore  an extended star formation history \citep{Clark09}.  No directed search has been done in W51 to identify  the lower-mass stars formed in the same star formation epoch as LS1. The spatially variable extinction and diffuse emission makes it hard to obtain accurate ages with only photometric data. 
Some of the  observed B stars above the  isochrones could be part of this population, even though the B stars are not located in the same field as LS1 (Field I), but in fields II and III.  


We know from the longer wavelength observations that currently star formation is still active, especially in the W51A cloud \citep{Ginsburg15}. In G49.5-04 two proto clusters are still forming \citep{Ginsburg12}. In the near-infrared we are sensitive to find stars with an age of a few Myrs, while at mm wavelength, where extinction is not an issue, much younger objects can be identified. However, the near-infrared traces the only spectral window to directly detect and characterize the photosphere of the most recently formed OB stars. 

Concluding, we found a very young massive star population consistent with an age of 3 Myr or less. Other epochs of star formation in W51 are found via mm observations (the ongoing star formation) and the identification of LS1, an evolved massive star. The evolutionary state of some of the B stars remains inconclusive as they could belong to either  older or the very young stellar population.


\begin{table*}[!t]
\caption{Properties of different \hii\ regions and their massive stellar content}
\label{regiontable}
\centering
\begin{tabular}{r r r r r r r}
\hline\hline
\hii\ region & Field&  Molecular mass\tablefootmark{a} &   Stellar mass & SFE& Most massive star  & Massive stars \\
	        & &     (\msun)         &     (\msun)        &     \%   &   & \#      \\
\hline
Total          &  & $1.4\times10^6$\tablefootmark{b}  &             $\sim4 \times 10^4$  & 3& 57: $\sim$120 \msun   & \\
G49.58-00.38 & I&	 \tablefootmark{c}                 & 	3100	\tablefootmark{d}		   & --- & LS1: $\sim$ 25 \msun	 &	LS1\\
G49.5-0.4  & II& $8.7 \times 10^4$    & 17600\tablefootmark{d} & 17 & 57: $\sim$120 \msun & 3, 6, 7, 12, 13, 16, 50, 57, 61, W51d\\
G49.2-0.3 &  III & $3.8\times10^4$      & 8000\tablefootmark{e}   &17 &2: $\sim$68 \msun & 2, 4, 5, 10, 14, 15\\
 G48.9-0.3 & IV &      $1.9\times10^4$                             &10000 \tablefootmark{e}& 35  &1: $\sim$50 \msun& 1, 8, 9, 11 \\
 \hline
\end{tabular}
\tablefoot{
\tablefoottext{a}{\citet{Kang10}}
\tablefoottext{b}{\citet{Carpenter98}}
\tablefoottext{c}{G49.58-00.38 is part of the G49.5-0.4 complex}
\tablefoottext{d}{\citet{Okumura00}}
\tablefoottext{e}{\citet{Kumar04}}
\label{tab:sfes}
}
\end{table*}

\subsection{Cluster properties}\label{sec:discussion_clusterprop}

The properties of the clusters inside the \hii\ regions where we have identified OB stars are summarised in Tab. \ref{tab:sfes}. The molecular masses are derived based on CO observations by \citet{Kang10} for the W51A cloud and by \citet{Carpenter98} for the \hii\ regions in W51B.  The stellar masses are estimated by \citet[W51a]{Okumura00} and \citet[W51B]{Kumar04} based on near-infrared photometry. \citet{Okumura00} derived a slope of the IMF of 1.8 (with Salpeter being 1.3), based on the more massive stars, which they extrapolated to lower masses to derive the total cluster mass. \citet{Kumar04} derived their cluster masses using the procedure outlined in \citet{Lada03} making use of the K-band luminosity function.

Based on the masses of the spectroscopically identified OB stars we can  get an independent estimate of the cluster mass. We assumed a \citet{Kroupa02} IMF and used the number of stars (eight) with photometric masses between 37 and 120 \msun\ identified by us in the W51 cloud as boundary condition. We carried out Monte Carlo simulations by randomly drawing a population from the IMF as described in \citet{Brandner08}. Based on this we derived a most likely cluster mass of $5 \times 10^3$ \msun, a factor of ten lower than the literature value based on near-infrared imaging \citep{Okumura00,Kumar04}. This difference shows how much we are incomplete in sampling the high-mass stellar population. As can be seen the in CMD  (Fig. \ref{fig:cmd}) and CCD (Fig. \ref{fig:ccd}) of W51, there are many bright, reddened point sources from which we do not have near-infrared spectra.  Even the near-infrared photometry mass estimate is likely still a lower limit, especially in the very extincted regions like G49.5-0.4. 

Once we know the molecular and stellar masses we can estimate the star formation efficiency (SFE). Using the cluster masses from  \citet{Okumura00} and \citet{Kumar04} and the molecular mass of \citet{Carpenter98}  we derive a SFE of 3 \% for the entire W51 complex.  This is similar to what is found on GMC scales in the Milky Way as well as other galaxies \citep[e.g.][]{Murray11,Kennicutt12}. Based on the YSOs detected in Spitzer images by \citet{Kang09}, \citet{Kang10} derived a SFE of $\sim$1 \% for the entire W51 region.

When we zoom in to the individual \hii\ regions, we find much higher SFEs. For G49.5-0.4 (our field II) a SFE of 17 \% is found based on the near-infrared data of \citet{Kang10} and the molecular observations of \citet{Carpenter98}. This is likely still even a lower limit as the stellar population is so reddened that even in the near-infrared a large fraction will be missed. In fact, all the spectroscopically detected O stars in G49.5-0.4  have \ak\ $>$ 2.5 mag. For the two regions in W51B we find 17 \% for G49.2-0.3 (field III) and even 35 \% for G48.9-0.3 (field IV). Derivation of SFEs towards the denser cores of GMCs in which the clusters are forming are generally much higher than SFEs averaged over an entire GMC. Simulations predict that the gravitationally bound regions of the clouds form star clusters with a higher SFE than the more dispersed population formed in unbound regions of the molecular clouds \citep{Bonnell11}. Similar results have been found in starburst region W43 where the SFE increases with increasing density of H$_{2}$ \citep{Louvet14}.  

On the other hand, gas dispersion due to stellar feedback can also result in a measured high star formation efficiency.  \citet{Ginsburg15} shows that in W51B the star formation has mostly stopped and there is little dense gas left. The feedback of the massive stars in these two clusters is now dispersing and/or ionizing the cloud. This could be the reason for the high SFE in G48.9-0.3, where already a significant fraction of the molecular material is removed. The cluster contains an early O star (\#1) and the ionizing photons and stellar wind emitted by the star is efficient in removing, especially, the low-density molecular gas \citep{Dale13}.

%
%
%
%
%

\subsection{Spatial distribution of the massive stars in GMC region}

Using Fig. \ref{fig:W51subregion} we can look at the spatial distribution of the (massive) stars and compare this to other star-forming complexes in the literature and relate the spatial distribution to the formation mechanisms of OB associations and star clusters. The W51 GMC contains many clusters spatially separated from each other, each surrounded by their own \hii\ region (e.g. G48.9-0.3 and G49.2-0.3). When we compare the location of the spectroscopically identified massive stars we find that not all the massive stars are located inside the clusters.  In both  G48.9-0.3 and G49.2-0.3 we found massive stars in the central region, however, there are also several B stars dispersed around the cluster. (\#8 in G48.9-0.2 and \#4, \#14 and \#15 in G49.3-0.4, Fig. \ref{fig:W51subregion}). In these regions, at least the massive stars are located where the diffuse  emission (\brg)  from the \hii\ region is located. The B stars do not emit enough ionizing photons to create a detectable \hii\ region. 

Most massive stars in our sample are located in  G49.5-0.4 (field II), the most massive region in W51. Fig. \ref{fig:W51subregion} panel III shows their spatial distribution.  We found only  few O stars associated with the currently forming proto-clusters associated with  the \hii\ regions W51d and W51e. These stars (W51d and IRS2E) are very deeply embedded and we clearly miss a large fraction of the stars forming there. We also found a several of the spectroscopically identified massive stars outside these two clusters. Some of the 
O stars we detected (\#6, \#7) are associated with diffuse emission likely caused by  \brg\ emission from a small \hii\ region (see also Sect. \ref{sec:radio}). Similar results have been found for the youngest massive stars in W51, where \citet{Saral17} found that most massive young stellar objects are associated with  \hii\ regions.  However, we also find sources not associated with \hii\ regions: sources 57 and 61 from \citet{Figueredo08} are not (anymore) associated with any \hii\ regions, but are among the most massive stars detected in this region (Tab. \ref{tab:logl_logt}). 

The observation that not all massive stars in G49.5-0.4 are part of the two forming proto-clusters  can be explained by two scenarios. First, the stars outside the cluster can be run-away stars \citep{Blaauw61}, where the star is ejected out of the star cluster by the supernova explosion of the binary companion or via dynamical interactions in the cluster \citep[e.g.][]{Gvaramadze09}. Taking a typical velocity of 100 \kms\ \citep{Gvaramadze10} and assuming an age of 2 Myrs would result at a displacement of $\sim$200 pc away from the birthplace. This distance is larger than  the size of the W51 GMC, indicating that these stars could be runaway stars. As the stars are still very young, the supernova kick scenario would be unlikely. In the dynamical ejection scenario, the less massive star involved in the dynamical three- or four body interaction is expelled  \citep[e.g.][]{Gvaramadze09}. The two stars 57 and 61 are among the most massive stars in W51, making also this scenario not very likely. 

The other explanation is that  massive stars do not exclusively form in the highest density regions of the GMC where the proto-clusters are forming, but also in their immediate surroundings.  Observations of other large star forming sites such as G305 \citep{Davies12}, the Carina region \citep{Smith06,Feigelson11}  and 30 Doradus \citep{Bressert12} showed similar behaviour. These regions contain dense star clusters but also a   spatially extended O star population. In W49 we found that  massive stars are more centrally concentrated around the main cluster \citep{Wu16}, however also there we identified several stars outside the central cluster.

This spatial distribution can be related to the properties of the GMCs in which the stars have formed. GMCs do not have a smooth density distribution, but are highly turbulent resulting in a hierarchical distribution of densities.  Modelling by \citet{Bonnell11} shows that the densest, gravitationally bound regions in the GMCs could give rise to gravitationally bound stellar clusters formed with a high star formation efficiency, while in the unbound parts of the clouds a more spatially dispersed population forms.

\begin{figure}[!t]
   \includegraphics[width=\hsize]{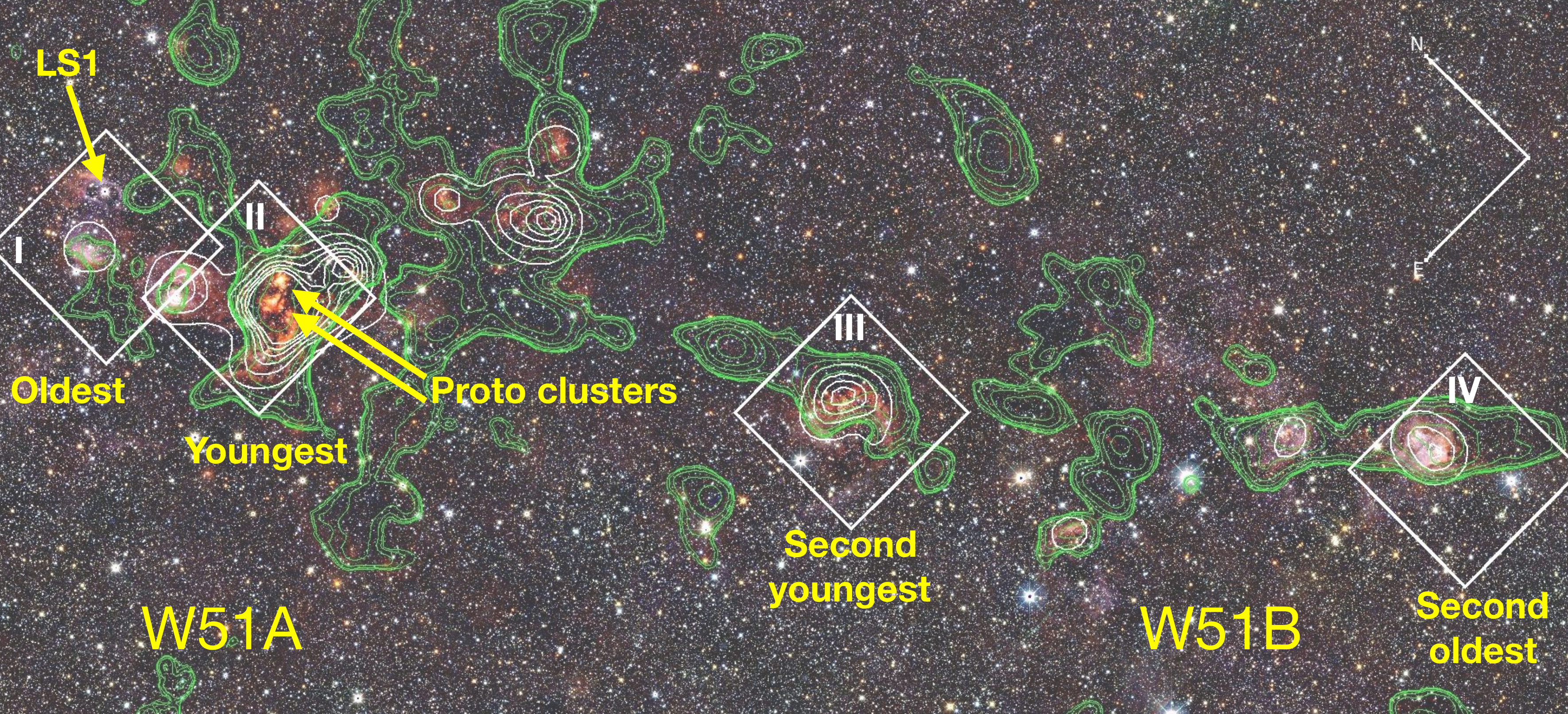}
      \caption{Three-colour image of W51 (identical to Fig. \ref{fig:largescale}) overlayed with the annotation of the evolutionary state of the different clusters discussed in Sect. \ref{sec:sfh} as well as the location of the two proto clusters identified in \citet{Ginsburg12} and the evolved massive star LS1. }
    \label{fig:sketch}
\end{figure}

\subsection{Star formation history}\label{sec:sfh}

The different stellar populations discussed in this paper and many other studies, show that W51 has a complex star forming history.
We find a population of massive OB stars with an estimated age of 3 Myrs or less. The presence of the evolved P Cygni-type supergiant suggests that star formation at some rate has been going on for longer than 3 Myrs  \citep{Okumura00,Clark09}.  Additionally, in the centre of W51A, two proto-clusters \citep{Ginsburg12} are located which are currently forming stars on a very high rate.  This raises the question how the star formation was initiated and how it proceeds. Due to the large extend and elongated shape of the W51 GMC several authors suggested, or could not exclude the possibility, that an external triggering mechanism working on large scales, such as a spiral arm passage is responsible for the star formation \citep{Koo99,Kumar04,Clark09}.

On the presence of smaller scale triggering, however, the opinions highly vary.  \citet{Koo97} found evidence for the presence of shockwaves in the interface between the supernova remnant  W51C and the GMC from HI and X-ray observations, suggesting that the supernova has affected the molecular cloud. However, considering the estimated age of the supernova remnant W51C of only 30 kyr \citep{Koo95}, the formation of the stars we observed in G49.2-0.3, the \hii\ region closest to W51C,  could not be triggered by the supernova event. This is supported by the lack of dense gas  showing that the collect-and-collapse scenario is not at play in this region \citep{Ginsburg15}. 

Sequential star formation in G49.5-0.4 was proposed by \citet{Okumura00}. By investigating their NIR photometry, an age spread between their regions 1, 2 and 3 has been found, implying the internal triggered star formation from north to south. \citet{Clark09} object against this hypothesis because if region 1 triggered similar star formation activity in region 3, then we might expect to see a large wind blown cavity in mid-IR created by stars like LS1 in region 1. But  Spitzer observations reveal that IRS2 does not resides on the boundaries of any wind bubble, implying IRS2 formed independently of the effect of other regions.  Additionally, evidence for cloud-cloud collisions inside G49.5-0.4  has been presented by \citet{Kang10} and \citet{Fujita17}.

Based on our spectroscopy in combination with the other literature data we can derive a relative time line for the four regions we have studied in detail. The \hii\ region in field I, G49.58-0.038 contains little dust and ionized gas as traced by  the green and white contours in Fig. \ref{fig:largescale}.  This suggests that this region is one of the older regions in W51 as most of the gas and dust have been dispersed. Additionally, towards this region we observe also the evolved star LS1. Under the assumption that this star is associated with the HII region, this would place an age estimate of  3 - 6 Myr \citep{Clark09} on this region.

For the other regions, the HRD analysis shows that they all have an age younger than $\sim$3 Myr. However, based on additional, more indirect arguments we can still derive a relative sequence. The youngest of the regions is vey likely G49.5-0.4 (field II), because of its high gas density and the presence of two proto-clusters. The extinction towards the massive stars in this region is the highest of all 4 regions studied. 

From the two regions in W51B, G48.9-0.3 (field IV) is likely older than G49.2-0.3 (field III). The OB stars identified in G48.9-0.3 have the lowest extinction of all the four regions, suggesting that the cluster has cleared out most of the dust. Additionally, the amount of molecular material around this cluster is also relatively low, resulting in the very high SFE derived for this region (Sect. \ref{sec:discussion_clusterprop}). The cluster is already relatively evolved and has dispersed a large fraction of its parental molecular cloud. This results in an evolutionary sequence (see Fig. \ref{fig:sketch}) with G49.58-0.038 being the oldest, after that G48.9-0.3 formed, then G49.2-0.3, and finally G49.5-0.4, which is currently still forming stars at a high rate.

Their spatial distribution excludes the triggering scenario proposed by \citet{Okumura00} as the two oldest regions are at the two extreme ends of the molecular cloud. This sequence confirms the conclusions of \citet{Clark09}, suggesting that star formation in W51 is a multi-seeded process where molecular clumps in the W51 GMC collapse at slightly different times resulting in the observed star formation. In the densest regions, like G49.5-0.4 internal triggering likely plays a role in the formation of some of the stars, but overall, triggering is not the dominant mode of star formation in W51. 

Recent kinematic studies of OB associations showed that they are not expanding clusters \citep[e.g.][]{Baumgardt07}, but  are made of kinematically independent sub-structure. They consist of the product of low-density multi-seeded star formation where each of the sub-clusters  leaves their own kinematic signature \citep{Wright16,Wright18,Ward18,Kuhn18}. The observed properties of W51 show that  W51 is also such an OB association, where we currently witness several of the sub-clusters being formed, while others are already slightly more evolved.

\section{Conclusions}\label{sec:conclusions}
In this paper, we presented  $JH$\Ks\ imaging from NTT/SOFI and $K$-band spectroscopy from LBT/LUCI of the stellar content in the star-forming region of W51.  We derived the following results and conclusions on the W51 GMC:

\begin{enumerate}
\item{We constructed CMDs and CCDs of four fields for which we collected near-infared imaging and spectroscopic data.
We found that the locations of the OB stars in the CCD are consistent with reddened photospheres. Additionally, we found several sources showing an infrared excess, suggesting the presence of sources with circumstellar disks in W51. We found a range of extinctions, with the youngest region. G49.5-0.4 (field II) showing the highest values. G48.9-0.3 (field IV) shows very low extinction, suggesting that this region is already more evolved.}

\item{We identified 15 O and B stars based on their absorption line spectra  and one source with a circumstellar disk showing CO bandhead emission. The most massive star discovered by our spectroscopic survey is star \#3 (with an estimated mass of $\sim$100 \msun). We added four O stars identified in the literature and derived the age of the stellar population to be consistent with $\sim$3 Myrs or less.} 

\item{We analyzed the properties of the clusters in which we have identified OB stars and found star-formation efficiencies higher than the overall SFE of W51 (1-3\%). This is consistent to what is observed towards other GMCs. The more evolved cluster (G48.9-0.3) shows a very high apparent SFE (35\%). The SFE at the time of formation might have been substantially lower as the majority of the gas is already  expelled from the cluster.}

\item{Finally we did not find evidence for triggered star formation being the dominant mode of star formation in W51. We did not find a progressive age-space sequence in the GMC, as the two regions on the northern (field I) and southern end (field IV) seem to be the most evolved regions. We support the conclusions drawn in the literature that star formation in W51 is multi-seeded. We conclude that W51 is an OB association.}
\end{enumerate}

\begin{acknowledgements}
We thank the referee for the comments and suggestions which improved the paper.
We thank Kate Rubin, Jaron Kurk and Barry Rothberg for carrying out part of the observations.
A.B. acknowledges MPIA for hospitality and travel support. A.P. acknowledges support from Sonderforschungsbereich SFB 881 "The Milky Way System" (subproject B5) of the German Research Foundation (DFG). 
Based on data obtained from the ESO Science Archive Facility. 
This research has made use of NASA's Astrophysics Data System Bibliographic Services (ADS). This research has made use of the SIMBAD database, operated at CDS, Strasbourg, France \citep{Wenger00}. IRAF is distributed by the National Optical Astronomy Observatory, 
which is operated by the Associated Universities for Research in 
Astronomy, Inc., under cooperative agreement with the National Science 
Foundation.

\end{acknowledgements}

\bibliographystyle{aa}
\bibliography{W51_AB}

 \begin{appendix} 

\section{Discussion on individual sub-clusters in W51}\label{sec:subclusters}

This appendix provide a literature overview of the clusters identified in the four fields discussed in this paper.

\subsection{G49.58-0.038} 
This \hii\ region is one of the least embedded regions in W51 and located at the edge of the W51A cloud. The western region (Fig. \ref{fig:W51subregion}) is the host of one of the two very luminous stars (LS1) discovered by  \citet{Okumura00}.  A high resolution $K$-band spectrum of LS1 identifies it as a P Cygni supergiant \citep{Clark09} with an estimated age of 3 - 6 Myrs. There is no mm or radio continuum associated with this source (Fig. \ref{fig:largescale}), confirming the more evolved nature of this region.
The eastern nebula corresponds to the northern most \hii\ region inside G49.5-0.4 \citep[region h,][]{Mehringer94}. This region is still associated with dust continuum emission (and radio free-free emission) suggesting a  younger age. 

\subsection{G49.5-0.4}
G49.5-0.4 is the most well studied region and the most luminous region in W51 and comprises the  W51A cloud. This area is one of the most active star-forming region in our Galaxy and contains a very high dense gas fraction \citep{Ginsburg15}, indicative of high star formation activity.  The two main radio component of this regions are sources W51d and W51e \citep{Mehringer94},  associated with infrared sources IRS2 and IRS1, respectively \citep{Neugebauer69,WynnWilliams74} and  located in the southern part of our image of field II (Fig. \ref{fig:W51subregion}). These two infrared sources are in fact embedded clusters \citep{Kumar04}. There are two massive proto-clusters in this region, one associated with IRS2 and the other with the \hii\ regions e1 and e2 containing each more than 10$^4$\msun\ of gas and having high infrared luminosities \citep{Ginsburg12,Ginsburg17}.

Numerous studies have been performed  on W51d/IRS2, also called W51 North, using data from a wide wavelength range. Using mm continuum observations, \citet{Zapata09} reported the presence of a dusty circumstellar disk and outflow perpendicular to it in IRS2, suggesting a single very massive protostar with a central stellar mass of more than 60 \msun. The two bright sources IRS2E and IRS2W were identified from low resolution NIR images, and IRSE was suggested to be a small cluster of stars \citep{Goldader94}. High resolution imaging reveals that IRS2E as a single unresolved source \citep{Figueredo08}. The spectrum of IRS2E presented in \citet{Figueredo08} may reflect the nebular nature, when considering its similarity with the UC\hii\ region G25.2-1.74 \citep{Ostarspec05}. Mid-infrared (7.8-13.5\micron) observations resolved IRS 2 into seven sources, including four UC\hii\ regions and an embedded protostar \citep{Okamoto01}. Spectral types of the central ionizing sources derived from mid-infrared data are later than those from the radio continuum flux.

\subsection{G49.2-0.3}
\emph{G49.2-0.3} is the brightest \hii\ region in W51B. W51B is a long filamentary structure including three  additional UC\hii\ regions (G49.1-0.4, G49.0-0.3 and G48.9-0.3).  The dense gas fraction in W51B is much lower than in W51A \citep{Ginsburg15}, suggesting that most of the star  formation in W51B has taken place and more exposed clusters should be visible.

The four UC\hii\ regions are located almost in a line parallel to the Galactic plane \citep{Koo99} and are all associated with molecular gas. G49.2-0.3 is associated with the cloud NE, while the other three UC\hii\ regions are associated with the cloud SW according to their positions and velocities. The radio continuum morphology suggests G49.2-0.3 to be a ``blister'' type \hii\ region formed near the edge of a molecular cloud \citep{Koo99}.

The morphology of the G49.2-0.3 region is suggested to be the product of shock wave interaction between the supernova remnant W51C and the GMC \citep{Kumar04}. OH masers observed in the vicinity of G49.2-0.3 provide evidence for this interaction \citep{Brogan13}. To the south-east of G49.2-0.3 the pulsar wind nebula candidate CXO J192318.5+1403035 is found \citep{Koo05,Aleksic12,Brogan13} to be associated with the supernova remnant. 

\subsection{G48.9-0.3}
\emph{G48.9-0.3} is the most southern \hii\ region in the W51B cloud and hosts two clusters \citep{Kumar04}.  Fig. \ref{fig:W51subregion} shows that both clusters are centrally concentrated, located 1.3 pc from each other.  This cluster is associated with the 68 \kms\ cloud \citep[e.g.][]{Kang10,Ginsburg15}. \citet{Kang10} places this \hii\ region and the neighbouring G49.0-0.3 inside the cloud.  \citet{Ginsburg15} place the two \hii\ regions in front of the 68 \kms\ cloud based on \element[][] [] [2]{H}\element{CO} absorption measurements.

 \section{HRD with a different extinction law}
 
 \begin{figure*}[!h]
   \includegraphics[width=\hsize]{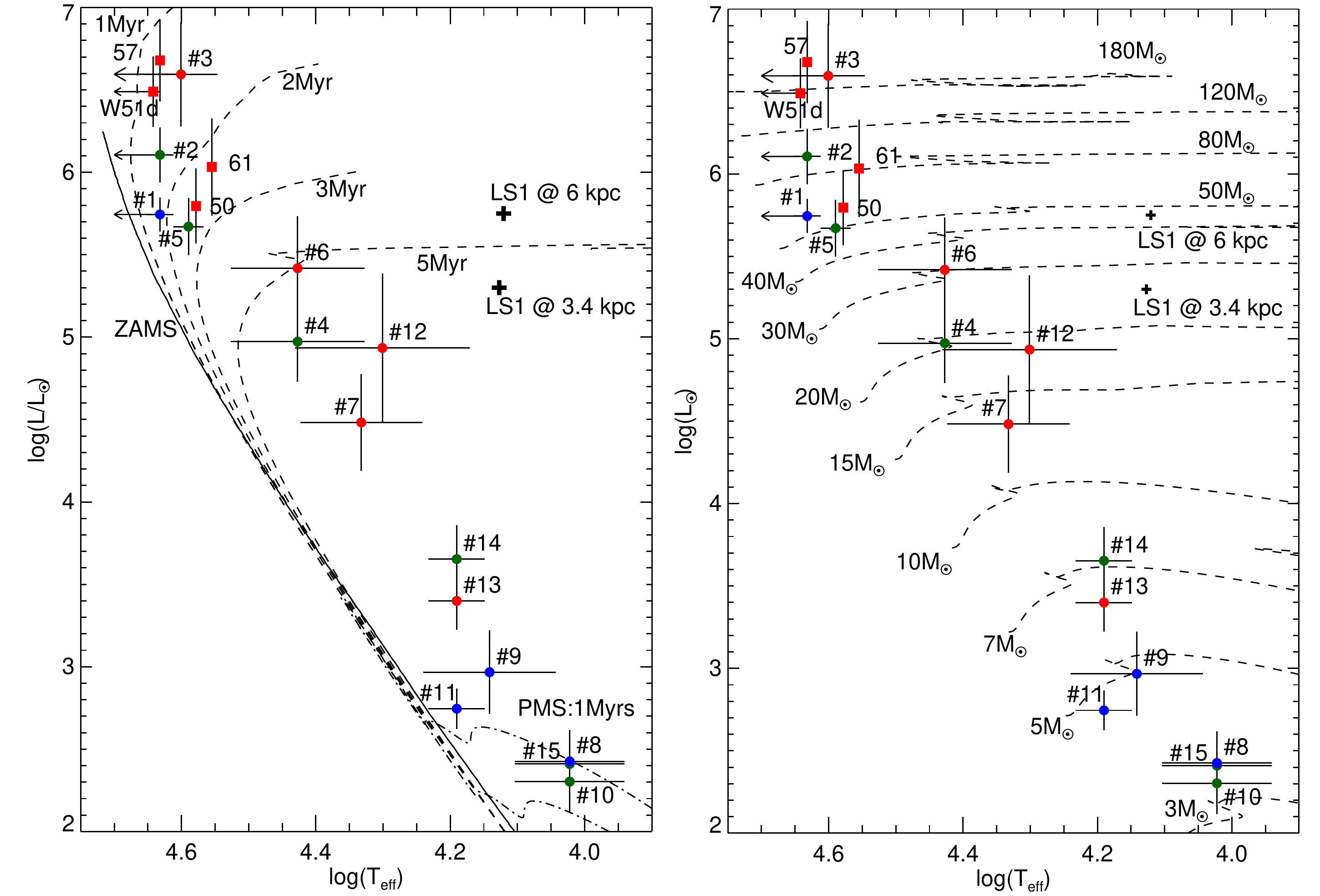}
      \caption{HRD of stars identified as OB stars in our spectroscopic survey. The coloured sources are the locations with the extinction correction of \citet{Indebetouw05}. The colour coding of the sources indicates in which region they are located: red: field II, green field III and blue field IV, the location of LS1 \citep{Okumura00} for the assumed distances of 3.4 and 6 kpc is marked with a "plus" sign (field I).  The red squares in panel d are the massive stars classified by \citet{Figueredo08} and \citet{Barbosa08}. 
      \emph{Left:} The solid line represents the ZAMS isochrone from \citet{Lejeune01} and the dashed lines the main-sequence isochrones for 1, 2, 3 and 5 Myr from \citet{Ekstrom12} and \citet{Yusof13}. The dash-dot lines represent the PMS isochrones of 1 and 2 Myr from \citet{Tognelli11}.  \emph{Right:} The dashed  lines represent the evolutionary tracks from \citet{Ekstrom12} and \citet{Yusof13}. 
      }
    \label{fig:HRD_Ind}
\end{figure*}

 \section{Late type stars towards W51}

\begin{table*}
\caption{Spectroscopically classified  late type stars towards W51}
\label{tab:LTstars}
\centering
\begin{tabular}{l  c c l l l }
\hline\hline
 ID & Ra & Dec & J & H & K   \\
              & ( h m s)       &   (\degr\ \arcmin\ \arcsec)      & (mag) & (mag) & (mag)  \\
\hline

OBS1MOS14   & 19 23 37.102   & +14 30 21.040 &   13.98     $\pm$ 0.01    &   12.77    $\pm$ 0.01   &    12.16    $\pm$ 0.01  \\
OBS1MOS15   & 19 23 32.933   & +14 31 52.800 &   14.80     $\pm$ 0.01    &   13.36    $\pm$ 0.01   &    12.70    $\pm$ 0.01\\
OBS2MOS07   & 19 23 50.448   & +14 32 57.460 &    9.55       $\pm$ 0.01  &     8.07      $\pm$ 0.01 &       7.56     $\pm$ 0.01\\
OBS2MOS08   & 19 23 50.822   & +14 33 16.910 &   14.10     $\pm$ 0.01    &   12.97    $\pm$ 0.01   &    12.42    $\pm$ 0.01\\
OBS2MOS09   & 19 23 50.876   & +14 33 34.550 &   12.69     $\pm$ 0.01    &   11.18    $\pm$ 0.01   &    10.48    $\pm$ 0.01\\
OBS2MOS10   & 19 23 47.640   & +14 33 21.040 &      ---                                  &          ---            &         ---                                            \\
OBS2MOS12   & 19 23 46.493   & +14 31 24.840 &   13.18     $\pm$ 0.01    &   12.02    $\pm$ 0.01   &    11.47    $\pm$ 0.01\\
OBS2MOS13   & 19 23 43.490   & +14 31 40.600 &   14.19     $\pm$ 0.01    &   13.01    $\pm$ 0.01   &    12.48    $\pm$ 0.01\\
OBS2MOS15   & 19 23 46.124   & +14 32 53.650 &   11.77     $\pm$ 0.01    &   10.40    $\pm$ 0.01   &    9.82      $\pm$ 0.01\\
OBS2MOS16   & 19 23 45.698   & +14 32 45.630 &   13.41     $\pm$ 0.01    &   11.59    $\pm$ 0.01   &    10.68    $\pm$ 0.01\\
OBS2MOS18   & 19 23 46.683   & +14 30 55.290 &                  ---         &            ---             &                 ---                       \\
OBS2MOS20   & 19 23 50.197   & +14 32 47.190 &   16.48     $\pm$ 0.02    &   12.70    $\pm$ 0.01   &    10.64    $\pm$ 0.01\\
OBS2MOS21   & 19 23 43.813   & +14 31 00.180 &   15.39     $\pm$ 0.01    &   14.29    $\pm$ 0.01   &    13.73    $\pm$ 0.02\\
OBS3MOS08   & 19 23 34.981   & +14 31 02.820 &   15.97     $\pm$ 0.01    &   14.30    $\pm$ 0.01   &    13.54    $\pm$ 0.02\\
OBS3MOS12   & 19 23 41.161   & +14 30 52.200 &   15.65     $\pm$ 0.01    &   14.42    $\pm$ 0.01   &    13.85    $\pm$ 0.01\\
OBS3MOS18   & 19 23 41.538   & +14 31 45.130 &  15.24     $\pm$ 0.01     &  14.12    $\pm$ 0.01    &   13.62    $\pm$ 0.01\\
OBS3MOS19   & 19 23 43.624   & +14 29 46.370 &   13.23     $\pm$ 0.01    &   12.44    $\pm$ 0.01   &    12.17    $\pm$ 0.01\\
OBS3MOS20   & 19 23 39.281   & +14 31 28.730 &   18.07     $\pm$ 0.05    &   15.54    $\pm$ 0.02   &    14.22    $\pm$ 0.01\\
OBS4MOS11   & 19 23 39.144   & +14 29 27.570 &   15.09     $\pm$ 0.01    &   13.62    $\pm$ 0.01   &    12.89    $\pm$ 0.01\\
OBS4MOS17   & 19 23 44.805   & +14 31 08.090 &   13.94     $\pm$ 0.03    &   12.82    $\pm$ 0.02   &    12.31    $\pm$ 0.02\\
OBS5MOS09   & 19 23 52.690   & +14 36 08.890 &   12.87     $\pm$ 0.01    &   11.90    $\pm$ 0.01   &    11.46    $\pm$ 0.01\\
OBS5MOS10   & 19 23 51.929   & +14 36 00.990 &                   ---        &          ---               &                   ---                     \\
OBS5MOS11   & 19 23 50.450   & +14 36 48.420 &   13.71     $\pm$ 0.01    &   12.54    $\pm$ 0.01   &    12.03    $\pm$ 0.01\\
OBS5MOS14   & 19 23 45.175   & +14 35 31.490 &   14.30     $\pm$ 0.01    &   13.30    $\pm$ 0.01   &    12.83    $\pm$ 0.01\\
OBS5MOS15   & 19 23 55.411   & +14 36 02.600 &   14.24     $\pm$ 0.01    &   13.04    $\pm$ 0.01   &    12.48    $\pm$ 0.01\\
OBS5MOS16   & 19 23 55.799   & +14 36 51.220 &   11.40     $\pm$ 0.01    &   10.26    $\pm$ 0.01   &    9.74      $\pm$ 0.01\\
OBS5MOS17   & 19 23 53.729   & +14 36 45.670 &   14.08     $\pm$ 0.01    &   12.90    $\pm$ 0.01   &    12.34    $\pm$ 0.01\\
OBS5MOS18   & 19 23 56.730   & +14 37 12.540 &   14.17     $\pm$ 0.01    &   13.41    $\pm$ 0.01   &    13.08    $\pm$ 0.01\\
OBS5MOS19   & 19 23 49.109   & +14 36 43.340 &   14.63     $\pm$ 0.01    &   12.47    $\pm$ 0.01   &    11.44    $\pm$ 0.01\\
OBS7MOS14   & 19 22 14.058   & +14 02 56.690 &   14.25     $\pm$ 0.01    &   13.16    $\pm$ 0.01   &    12.67    $\pm$ 0.01\\
OBS7MOS20   & 19 22 17.018   & +14 02 08.350 &   13.89     $\pm$ 0.01    &   12.78    $\pm$ 0.01   &    12.30    $\pm$ 0.01\\
OBS8MOS09   & 19 22 14.314   & +14 03 42.440 &                 ---          &         ---                &            ---                            \\
OBS8MOS14   & 19 22 13.229   & +14 02 51.790 &   15.40     $\pm$ 0.01    &   14.30    $\pm$ 0.01   &    13.80     $\pm$ 0.01\\
OBS8MOS16   & 19 22 14.543   & +14 02 20.390 &   18.20     $\pm$ 0.05    &   16.00    $\pm$ 0.02   &    14.61     $\pm$ 0.01\\
OBS9MOS10   & 19 23 03.218   & +14 16 24.100 &   17.40     $\pm$ 0.05    &   15.35    $\pm$ 0.03   &    14.24     $\pm$ 0.03\\
OBS9MOS11   & 19 23 01.926   & +14 15 20.320 &   17.35     $\pm$ 0.02    &   14.22    $\pm$ 0.01   &    12.58     $\pm$ 0.01\\
OBS9MOS13   & 19 23 03.346   & +14 16 02.730 &   18.27     $\pm$ 0.05    &   15.61    $\pm$ 0.01   &    14.21     $\pm$ 0.01\\
OBS9MOS16   & 19 23 07.275   & +14 15 56.260 &   18.48     $\pm$ 0.06    &   15.81    $\pm$ 0.01   &    14.31     $\pm$ 0.01\\
OBS9MOS19   & 19 23 05.161   & +14 14 42.460 &   18.63     $\pm$ 0.06    &   16.06    $\pm$ 0.02   &    14.67     $\pm$ 0.02\\
OBS9MOS20   & 19 23 02.451   & +14 17 14.910 &   12.68     $\pm$ 0.01    &   10.32    $\pm$ 0.01   &    9.16       $\pm$ 0.01\\

\hline
\end{tabular}

\end{table*}

 \end{appendix} 

\end{document}